\documentclass[superscriptaddress,twocolumn,showpacs,preprintnumbers,amsmath,amssymb,floatfix,nofootinbib]{revtex4-1}
\usepackage{graphicx}
\usepackage{dcolumn}
\usepackage{bm}

\def\r{{\boldsymbol r}}
\def\R{{\boldsymbol R}}

\begin{document}

\def\bb    #1{\hbox{\boldmath${#1}$}}

\title{Glauber Monte Carlo predictions for ultra-relativistic collisions with ${}^{16}{\rm O}$}

\author{Maciej Rybczy\'nski}
\email{Maciej.Rybczynski@ujk.edu.pl}
\affiliation{Institute of Physics, Jan Kochanowski University, 25-406 Kielce, Poland}

\author{Wojciech Broniowski}
\email{Wojciech.Broniowski@ifj.edu.pl}
\affiliation{Institute of Physics, Jan Kochanowski University, 25-406 Kielce, Poland}
\affiliation{H. Niewodnicza\'nski Institute of Nuclear Physics PAN, 31-342 Cracow, Poland}

\date{26 November 2019}  

\begin{abstract}
We explore Glauber Monte Carlo predictions for the planned  ultra-relativistic  ${}^{16}{\rm O}$+${}^{16}{\rm O}$ and  p+${}^{16}{\rm O}$
collisions, as well as for collisions of  ${}^{16}{\rm O}$ on heavy targets. In particular, we present specific collective flow 
measures which are approximately independent on the hydrodynamic response of the system, such as the ratios of eccentricities obtained from cumulants with different
numbers of particles, or correlations of ellipticity and triangularity described by the normalized symmetric cumulants. 
We use the state-of-the-art correlated nuclear distributions for ${}^{16}{\rm O}$ and compare the results to the uncorrelated case, 
finding moderate effects for the most central collisions. 
We also consider the wounded quark model, which turns out to yield similar results to the wounded nucleon model for the considered measures. 
The purpose of our study is to prepare some ground for the upcoming experimental proposals, as well as to provide input for 
possible more detailed dynamical studies with hydrodynamics or transport codes.  
\end{abstract}


\maketitle

\section{Introduction \label{sec:intro}}

In the continued quest~\cite{Citron:2018lsq} for deeper understanding of the rich physics unveiled by ultra-relativistic nuclear collisions, proposals have been made 
to study collisions with ${}^{16}{\rm O}$ beams, both at the LHC  (see Sec.~9.10 of~\cite{Citron:2018lsq} 
for ${}^{16}{\rm O}$+${}^{16}{\rm O}$ and Sec.~11.3 or p+${}^{16}{\rm O}$, 
describing the experimental programs that could be carried out in runs beyond the year 2022)
and at RHIC~\cite{starO16}. 

Investigations of  ${}^{16}{\rm O}$+${}^{16}{\rm O}$ are motivated by the exploration of emergence of collectivity in small systems, which has
attracted a lot of attention over the past few years~\cite{Bozek:2011if,Kozlov:2014fqa,Bzdak:2014dia,Bozek:2013uha,Werner:2013tya,Nagle:2013lja,Bozek:2014cya,Bozek:2015qpa}.
This is a major issue, as it concerns the very nature of the initial dynamics in the created fireball (for recent overviews see, 
e.g.,~\cite{Gelis:2016upa,Busza:2018rrf,Mazeliauskas:2018wam} and references therein). 
A distinct feature of the ${}^{16}{\rm O}$+${}^{16}{\rm O}$ collisions is that with a similar number of participants as in the earlier 
studied  p+Pb collisions, the participants are distributed more sparsely in the transverse plane. This is expected to lead to 
different subsequent evolution.
 
Studies of p+${}^{16}{\rm O}$ collisions find a broader justification 
from the physics of air showers generated with cosmic rays~\cite{demb} and our lack of full understanding of the 
production process, e.g., the cosmic ray neutrino puzzle (see~\cite{Dembinski:2019uta} and references therein). 
They also carry significance for investigating the onset of collectivity in ultra-relativistic nuclear collision.

The purpose of this work is to provide some model predictions for the planned reactions that could be used in preparatory analyses for the experimental proposals. 
We use the Glauber~\cite{glauber1959high,Czyz:1969jg} modeling, which has become a basic tool to describe the 
initial state due to its simplicity and phenomenological success. Our simulations are carried out with {\tt GLISSANDO~3}~\cite{Bozek:2019wyr}.

We note that an analysis similar to ours has recently been carried out for the LHC energies by Sievert and Noronha-Hostler~\cite{Sievert:2019zjr}, 
where the {T\raisebox{-.5ex}{R}ENTo} code~\cite{Moreland:2014oya} has been used for the initial conditions and hydrodynamics run with 
v-USPhydro~\cite{Noronha-Hostler:2013gga}. 
Studies based on AMPT model~\cite{Zhang:1999bd} were presented by Huang, Chen, Jia, and Li in~\cite{Huang:2019tgz} for the RHIC collision energies.
The details of our model implementation 
concerning the distribution of nucleons in ${}^{16}$O as well as the NN reaction features are different from the above-mentioned approaches. Consequently, also 
the studied eccentricity measures differ to some extent, providing an independent estimate for model uncertainties in physical predictions for the 
considered reaction.

As the further evolution of the system with hydrodynamics or transport is outside of our present scope, we focus on flow observables 
which are not strongly sensitive to the hydrodynamic response, such as ratios of various flow coefficients or the normalized symmetric cumulants.

\section{Structure of ${}^{16}{\rm O}$ \label{sec:16o}}

Before we embark on collisions of ${}^{16}{\rm O}$, it is worth to focus on its nuclear structure. 
This is relevant, as properties of collisions reflect the features of the projectiles (as well as, of course, the 
NN collision mechanism). Needless to say, the size of the nuclei affect the total cross section, whereas 
two-body correlations influence to some degree the flow observables~\cite{Broniowski:2010jd}. 
To have the possibly most realistic ${}^{16}{\rm O}$ nucleus, rather than using parameterizations of its one-body distribution, we 
take configurations from  state-of-the-art dynamical nuclear physics calculations.
Specifically, we use 6000 configurations from 
cluster Variational Monte Carlo (CVMC) simulations~\cite{Lonardoni:2017egu} with the Argonne~v18 two-nucleon and Urbana~X three-nucleon potentials, as provided 
in files in~\cite{Loizides:2014vua}. We stress that these dynamically generated distributions contain realistic 
nuclear correlations, which are absent (or put in by hand in the form of 
a repulsive core for the two-body distributions) when some simple parameterizations of the nuclear one-body distributions are used.

\begin{figure}
\begin{center}
\includegraphics[width=0.4 \textwidth]{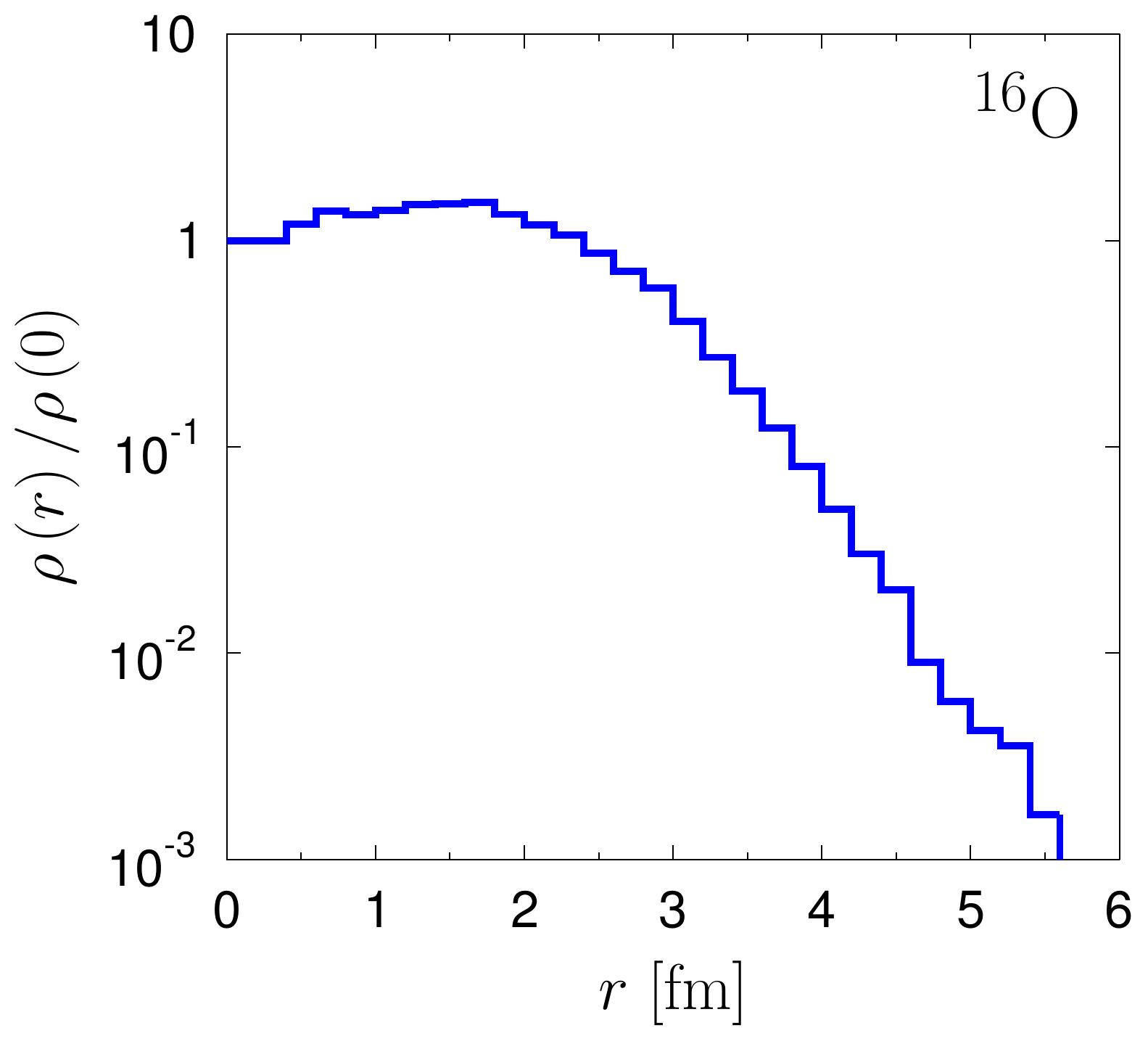} 
\end{center}
\vspace{-5mm}
\caption{Nuclear radial density, $\rho(r)$ (in units of the central density), of the ${}^{16}{\rm O}$ nucleus obtained from the 6000 configurations from the CVMC 
simulations~\cite{Lonardoni:2017egu}. \label{fig:dens}}
\end{figure} 

\begin{figure}
\begin{center}
\includegraphics[width=0.4 \textwidth]{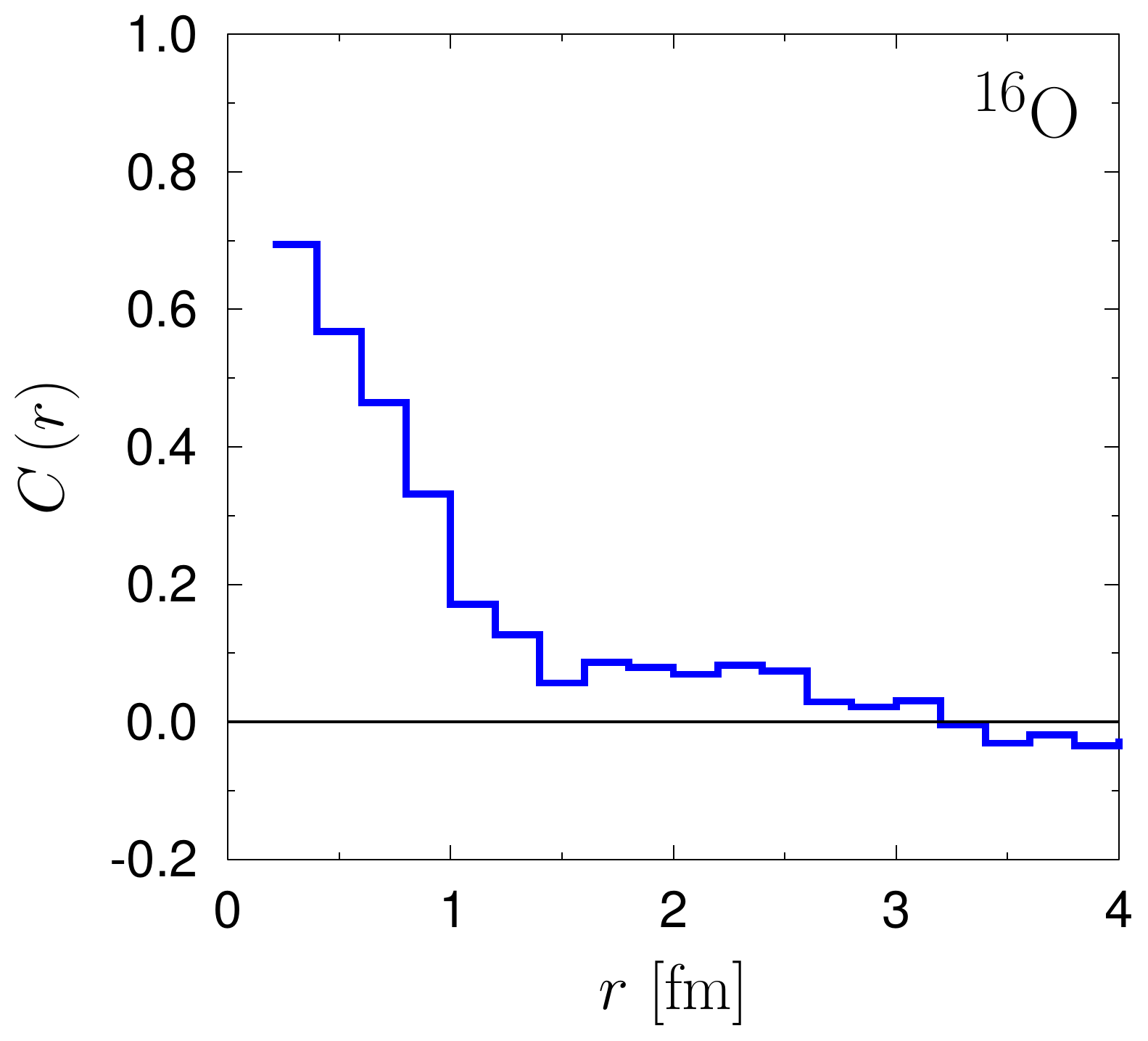} 
\end{center}
\vspace{-5mm}
\caption{Same as in Fig. \ref{fig:dens} but for the correlation function $C(r)$ defined in Eq.~(\ref{eq:corr}). \label{fig:corr}}
\end{figure}

The one-body nuclear radial densities are given in Fig.~\ref{fig:dens}, where we plot the distribution of the centers of 
nucleons, $\rho(r)$, 
conventionally in units of the central density $\rho(0)$. The corresponding ms radius 
is $\langle r^2 \rangle = (2.6~{\rm fm})^2$, which folded with the proton charge form factor with ms radius $\langle r^2 \rangle_p = (0.84~{\rm fm})^2$ 
yields the ms  charge radius of  ${}^{16}{\rm O}$ of
\begin{eqnarray}
\langle r^2 \rangle_{\rm ch} =  \langle r^2 \rangle+\langle r^2 \rangle_p=(2.7~{\rm fm})^2, \label{eq:size}
\end{eqnarray}
which is comfortably in the right experimental range of $(2.699(5)~{\rm fm})^2$~\cite{ANGELI201369}.

A standard measure of the nuclear two-body correlations is provided by the ratio between of
normalized two-body probability distribution of nucleons in their relative distance $r$ and the
folding of their one-body distributions, 
\begin{equation}
\label{eq:corr} C(r)= 1 -\frac
{\int d^3 R\, f^{(2)}(\R+\frac{\r}{2},\R-\frac{\r}{2})}
{\int d^3 R\, f^{(1)}(\R+\frac{\r}{2})f^{(1)}(\R-\frac{\r}{2})}.
\end{equation}
In Monte Carlo simulations, this ratio is easily obtained by generating histograms in the relative distance between all nucleon pairs
from the same configuration, divided by the ``mixed'' histogram, where the nucleons in the pair come from different configurations.
By construction, the ``mixed'' two-body distribution exhibits no correlations and plays the role of the denominator in Eq.~(\ref{eq:corr}).
The result of this procedure, obtained with {\tt GLISSANDO~3}~\cite{Bozek:2019wyr}, is shown in Fig.~\ref{fig:corr}. We note a soft-core
behavior at low separations $r$, where $C(r)>0$ indicates repulsion. At larger $r$ the correlations disappear, as expected. 
The noise in the figure is caused by rather low available statistics (6000 configurations from~\cite{Loizides:2014vua}).

In some cases, we will show comparisons to the case where 
the correlations are removed by the mixing method. The procedure used here is as follows: we take a nucleus whose nucleon positions are 
represented with spherical coordinates and regenerate randomly the angular coordinates, while retaining the radius. That way the 
correlations are removed, whereas the radial density distributions are preserved. This is important, as we do not wish to change size 
of the projectiles, which directly relates to the collision cross section. An equivalent method, yielding essentially the same results, is to 
form ``mixed'' nuclei with each nucleon taken from a different ``physical'' nucleus.   

To summarize this section, in the following analysis of nuclear collisions we are going to use realistic 
dynamically generated configurations of  ${}^{16}{\rm O}$, 
reproducing the charge radius and involving proper two-body correlations.

\section{Glauber modeling \label{sec:glauber}}

\begin{figure}
\begin{center}
\includegraphics[width=0.43 \textwidth]{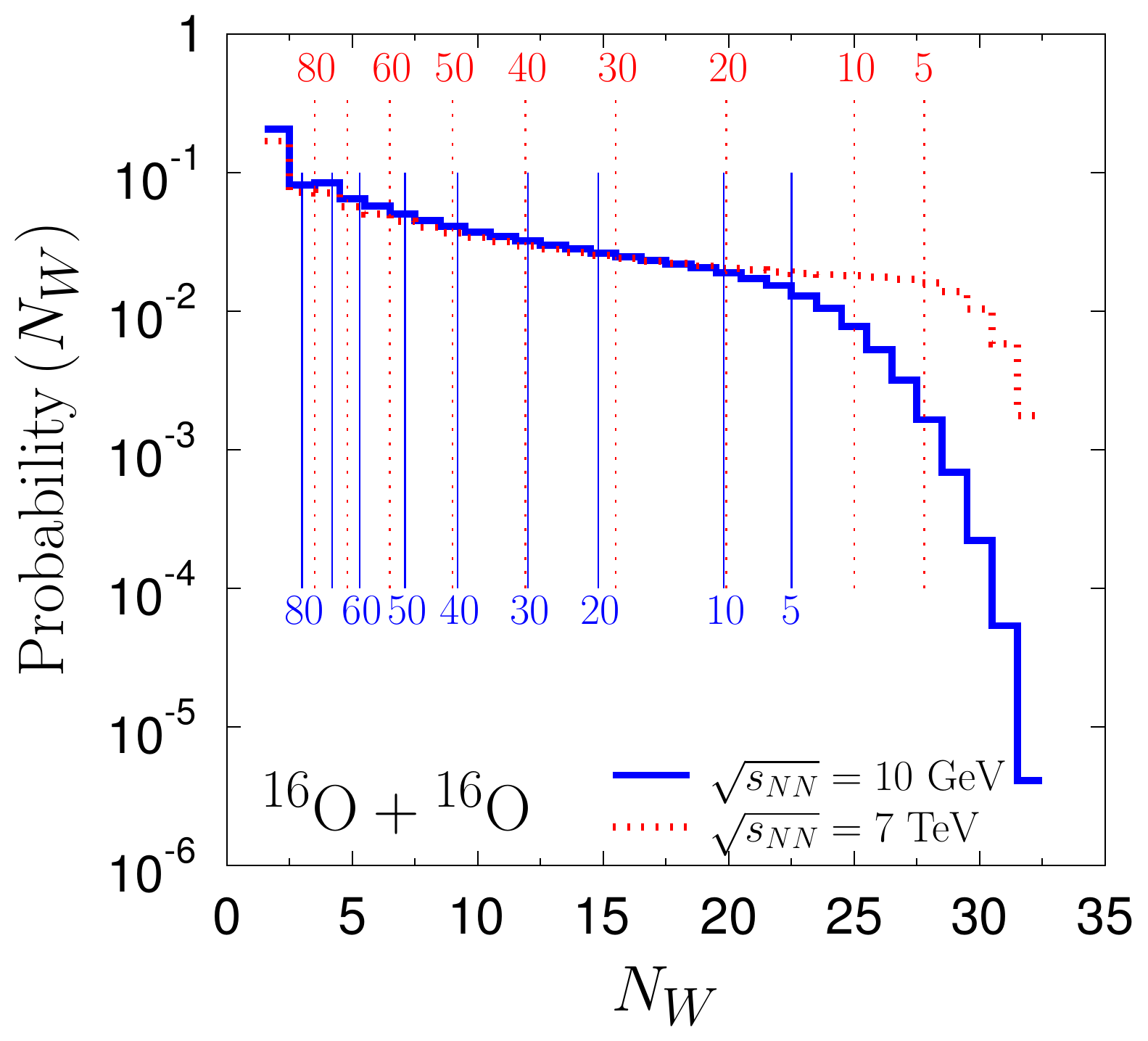} 
\end{center}
\vspace{-5mm}
\caption{Probability distributions of the number of wounded nucleons, $N_{\rm w}$, in ${}^{16}{\rm O}+{}^{16}{\rm O}$ collisions at $\sqrt{s_{NN}}=10$~GeV
(solid line) and 7~TeV (dashed line). The vertical lines indicate the boundaries of the corresponding centralities (in percent), 
with the lower labels corresponding to 10~GeV, and the upper labels to 7~TeV. \label{fig:nw}}
\end{figure} 

The applied Glauber Monte Carlo approach (for a review see~\cite{Miller:2007ri}) is described 
in detail in {\tt GLISSANDO~3}~\cite{Bozek:2019wyr}, so we are very brief here. Importantly, 
the adopted NN inelasticity profile, a.k.a. the Van Hove function~\cite{van1963phenomenological,VanHove:1964rp}, is obtained 
from fits of the COMPETE model parametrization implemented in the Particle Data Group review~\cite{Patrignani:2016xqp}. This parametrization provides the best 
available description of the $pp$ and $p\bar p$ scattering over the range of the collision energies from $\sqrt{s_{NN}}\simeq 5$~GeV up to the highest LHC energies.

\begin{figure*}
\begin{center}
\includegraphics[width=0.4 \textwidth]{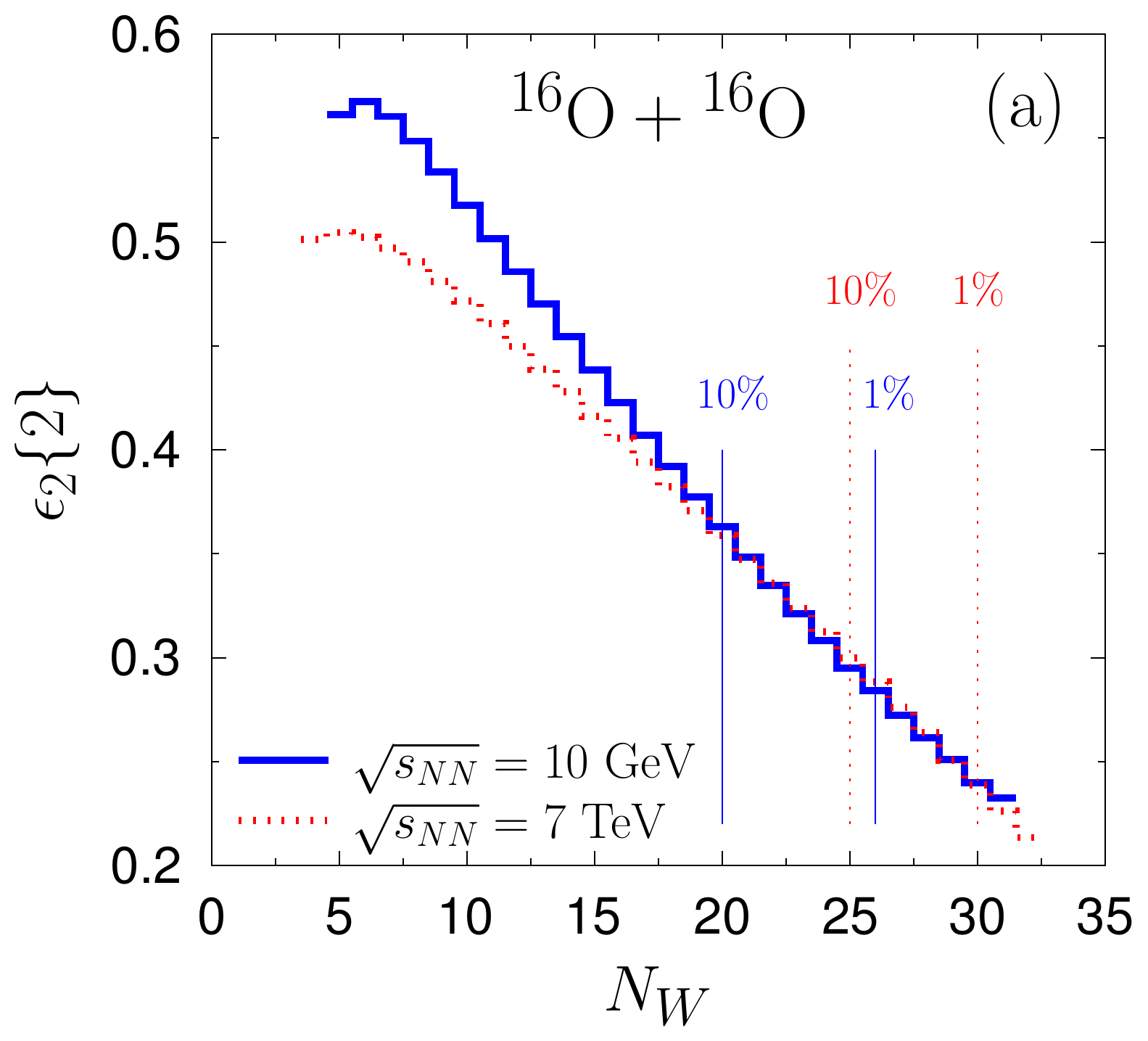} 
\includegraphics[width=0.4 \textwidth]{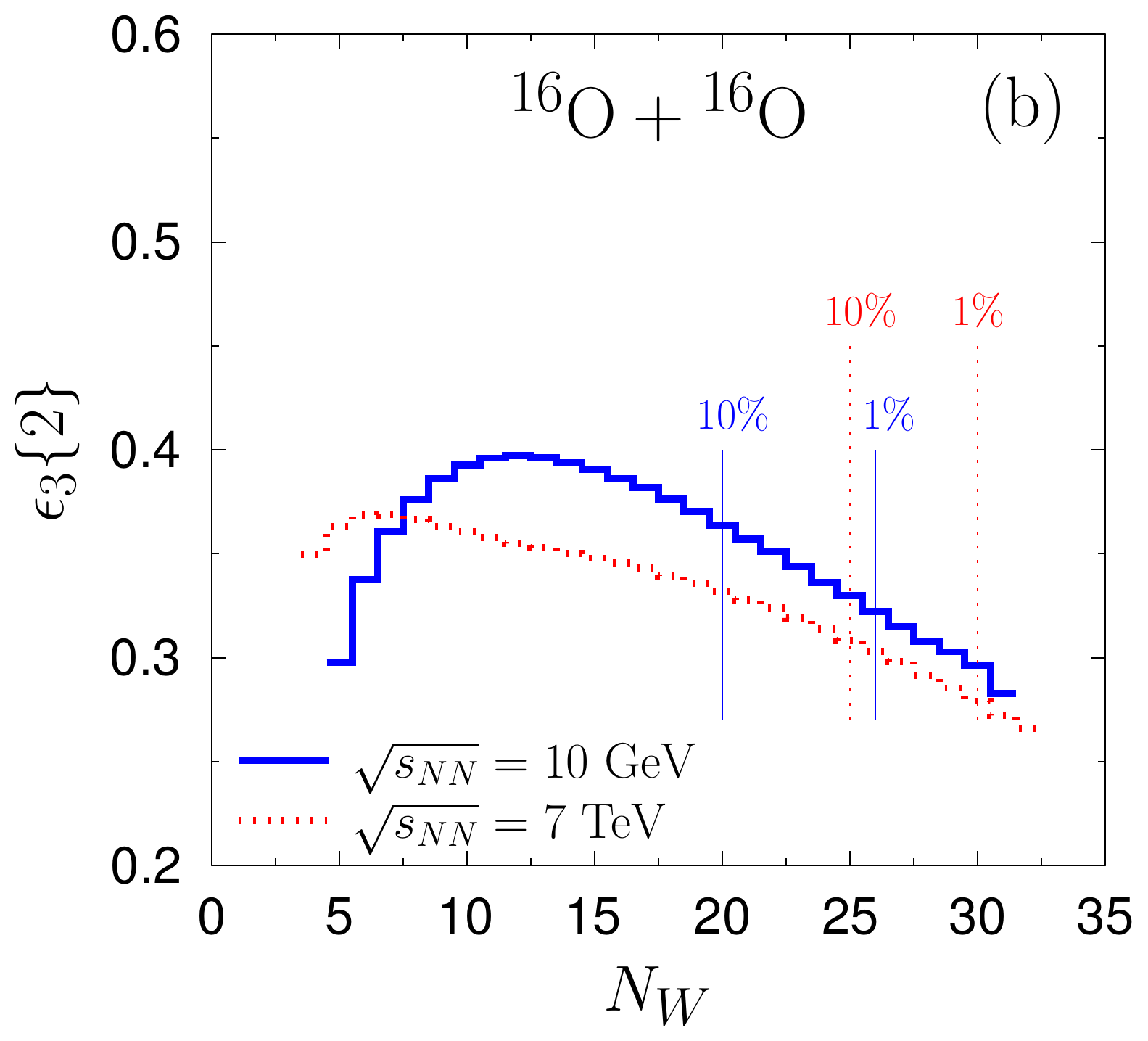} 
\end{center}
\vspace{-5mm}
\caption{Glauber model predictions for the ellipticity (a) and triangularity (b) obtained with two-particle 
cumulants for the fireball created in ${}^{16}{\rm O}+{}^{16}{\rm O}$ collisions at 
$\sqrt{s_{NN}}=10$~GeV and $\sqrt{s_{NN}}=7$~TeV and plotted as functions of the 
number of wounded nucleons, $N_w$. The vertical lines indicate the boundaries of the most 
central 1\% and 10\% classes for the two collision energies.\label{fig:eps_2}}
\end{figure*}

\begin{figure*}
\begin{center}
\includegraphics[width=0.4 \textwidth]{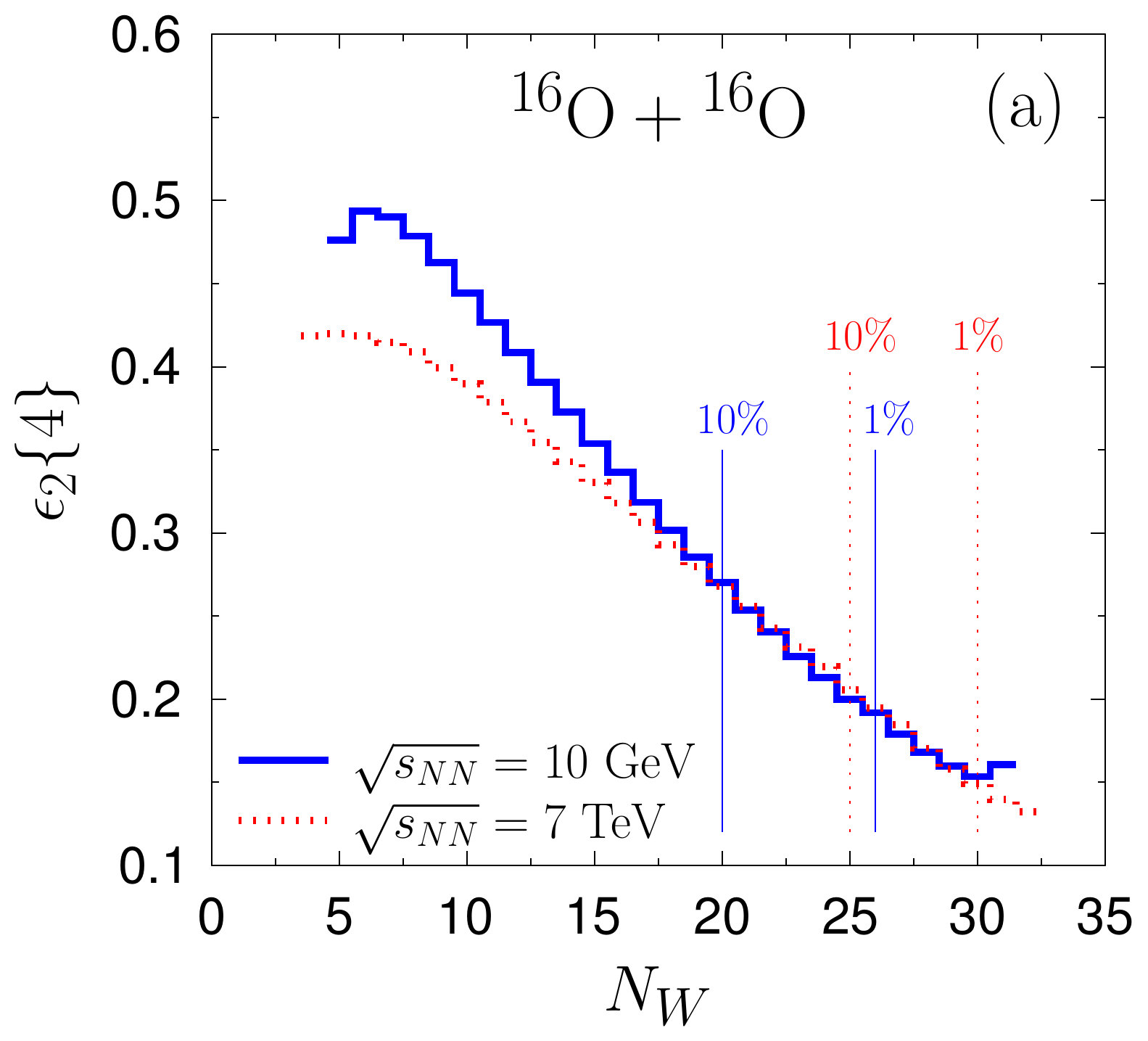} 
\includegraphics[width=0.4 \textwidth]{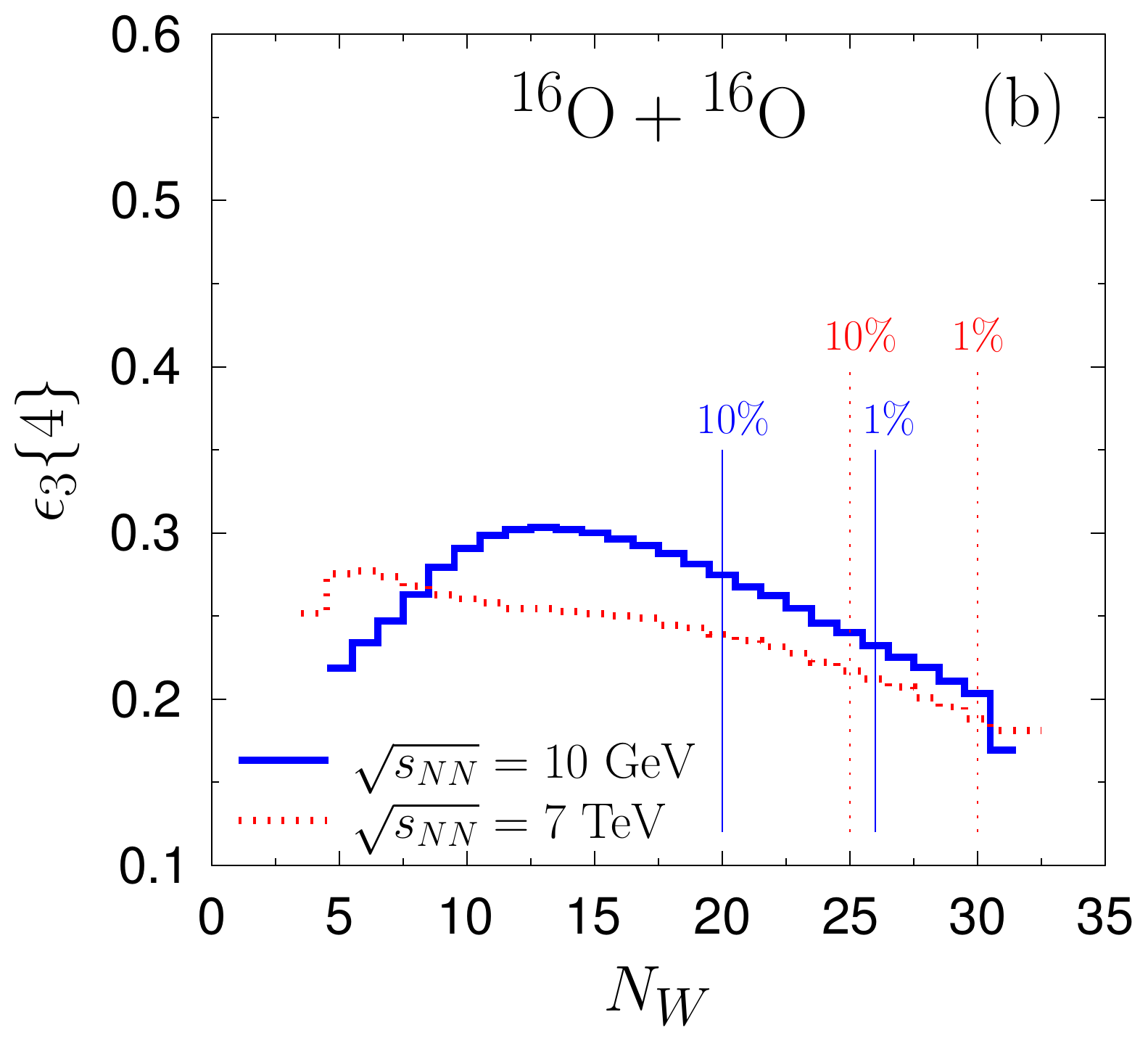} 
\end{center}
\vspace{-5mm}
\caption{Same as in Fig.~\ref{fig:eps_2} but for $\epsilon_n\{4\}$ from Eq.~(\ref{eq:ecc}). \label{fig:eps_4}}
\end{figure*} 

\begin{figure*}
\begin{center}
\includegraphics[width=0.4 \textwidth]{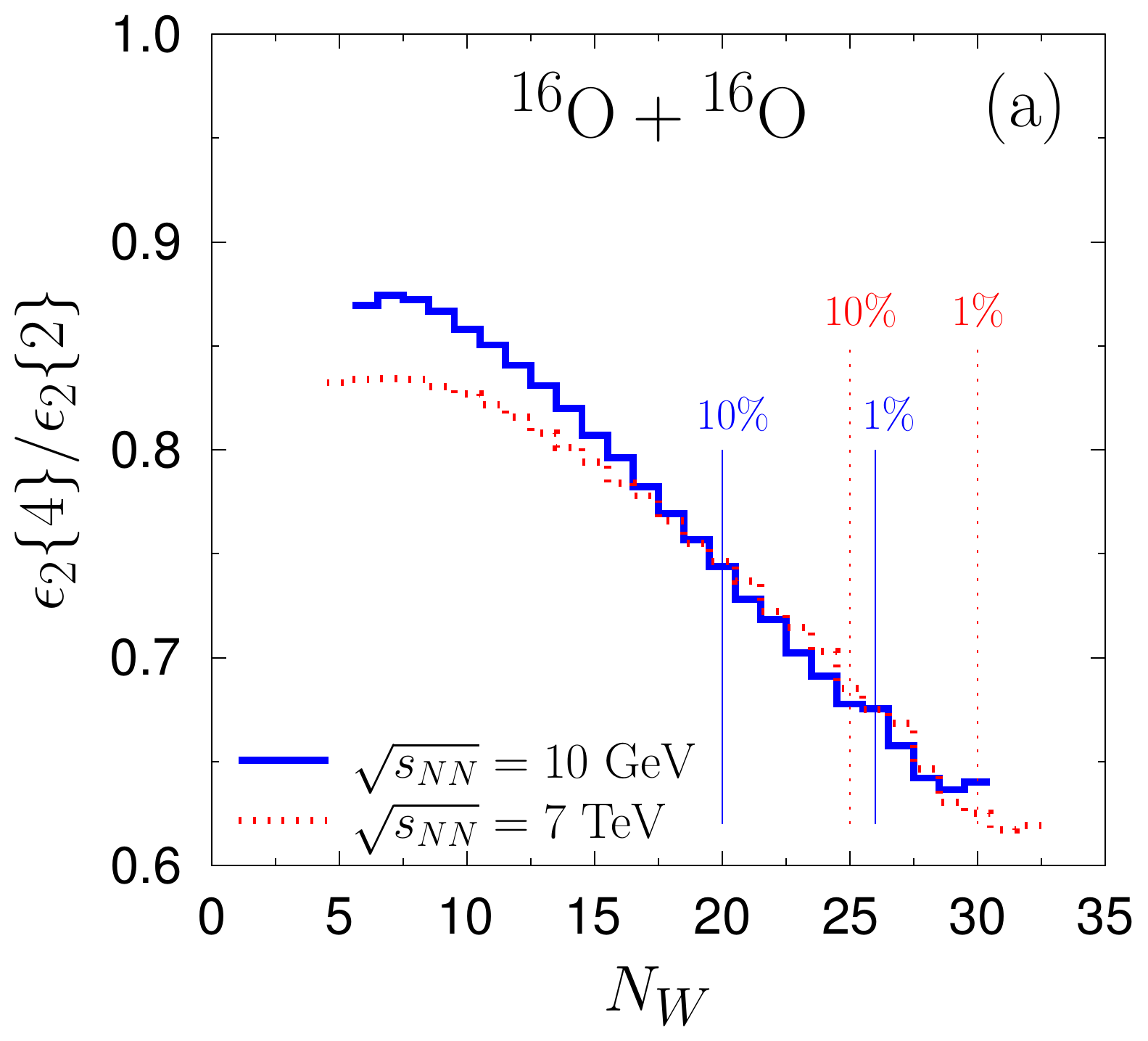} 
\includegraphics[width=0.4 \textwidth]{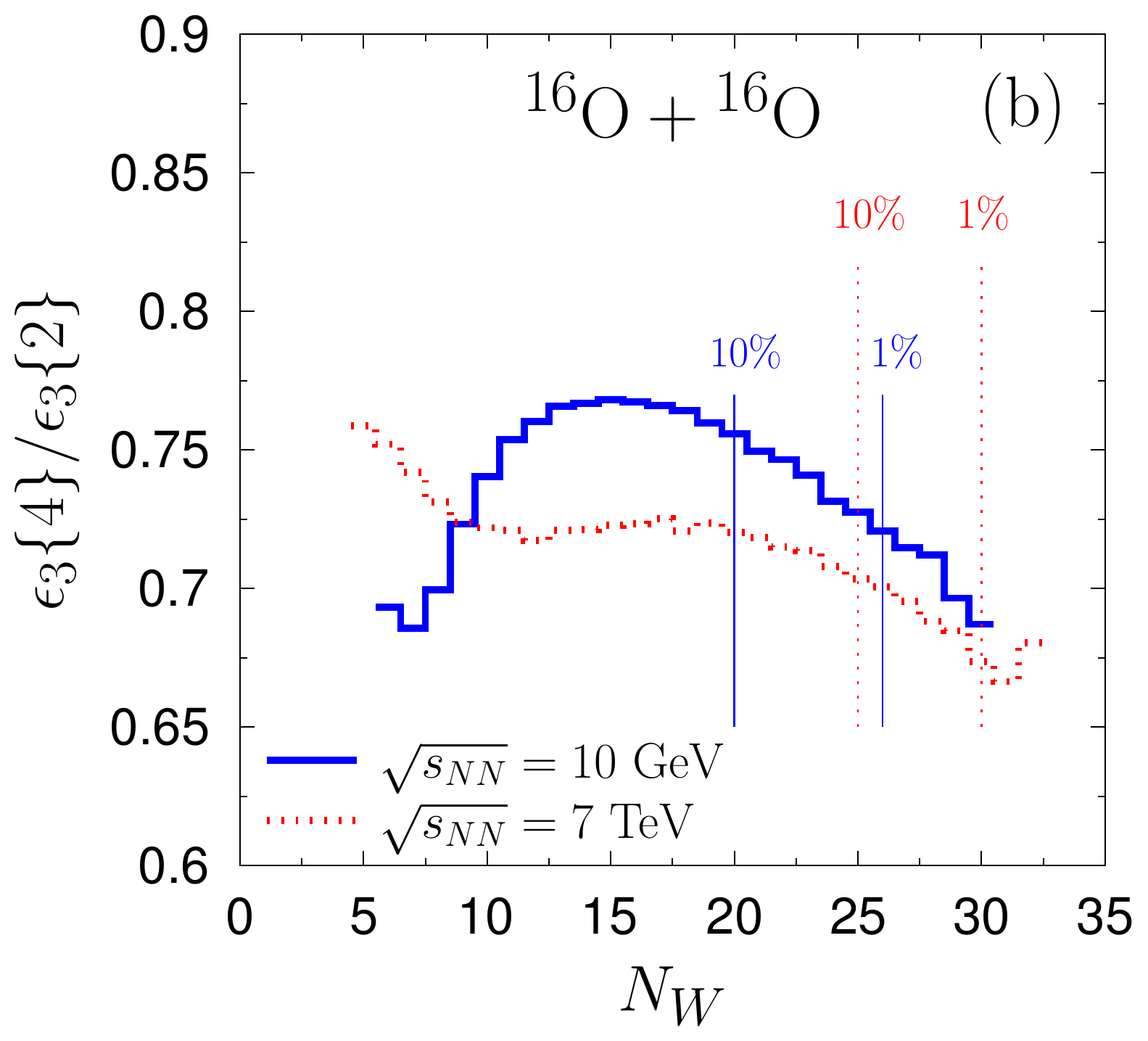} 
\end{center}
\vspace{-5mm}
\caption{Same as in Fig.~\ref{fig:eps_2} but for the ratio $\epsilon_n\{4\}/\epsilon_n\{2\}$ of Eq.~(\ref{eq:32}). \label{fig:eps_rat}}
\end{figure*} 

We use the wounded nucleon~\cite{Bialas:1976ed,Bialas:2008zza} variant of the Glauber model with an admixture of the  binary collisions~\cite{Kharzeev:2000ph},
which has been found necessary to describe the multiplicity distributions of the produced hadrons~\cite{Back:2001xy}. Thus the
initial entropy deposition in the transverse plane, $S(x,y)$, is proportional to a combination of the wounded and binary contributions controlled by 
the parameter $\alpha$, 
\begin{eqnarray}
S(x,y) \propto \frac{1-\alpha}{2} \rho_{\rm W}(x,y) +\alpha \rho_{\rm bin}(x,y), \label{eq:mixed}
\end{eqnarray}
where the densities $\rho_{\rm W}(x,y)$ and $\rho_{\rm bin}(x,y)$ are obtained from the positions of the point-like sources generated with the Glauber Monte Carlo, 
which are then smeared with Gaussian profiles of width 0.4~fm~\cite{Bozek:2019wyr}. As is well known, the smearing effect quenches the eccentricities of the 
fireball.
For the mixing parameter we take typical values, with $\alpha=0.12$ at $\sqrt{s_{NN}}=10$~GeV, and 
 $\alpha=0.15$ at $\sqrt{s_{NN}}=7$~TeV and 10~TeV.

\begin{figure}
\begin{center}
\includegraphics[width=0.4 \textwidth]{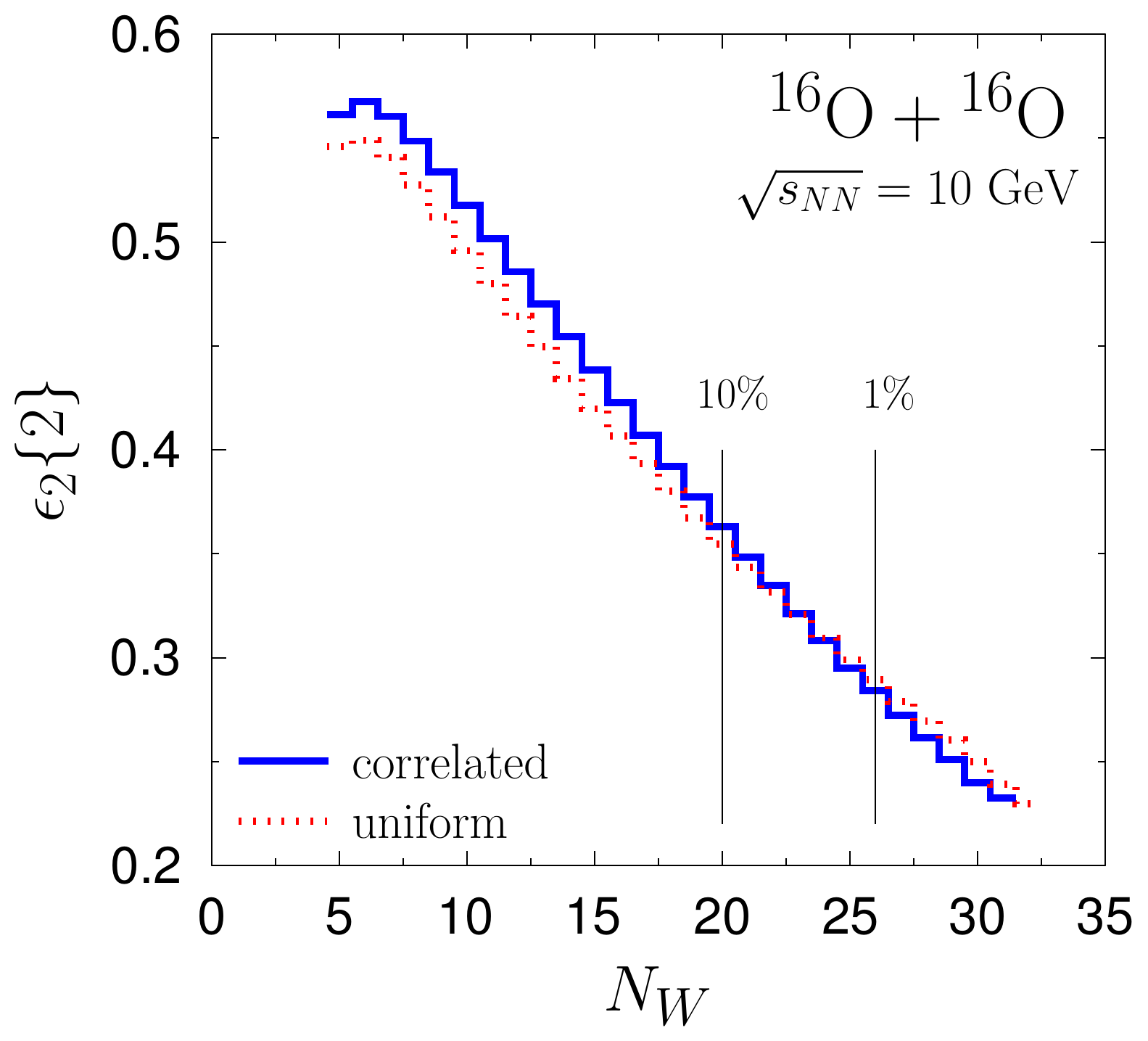}
\end{center}
\vspace{-5mm}
\caption{The eccentricity coefficient $\epsilon_2\{2\}$ for ${}^{16}{\rm O}+{}^{16}{\rm O}$ collisions at $\sqrt{s_{NN}}=10$~GeV  for correlated and uniform
${}^{16}{\rm O}$ nuclear distributions, 
plotted as functions of the number of wounded nucleons, $N_w$. \label{fig:eps_22_comp}}
\end{figure} 

\begin{figure}
\begin{center}
\includegraphics[width=0.4 \textwidth]{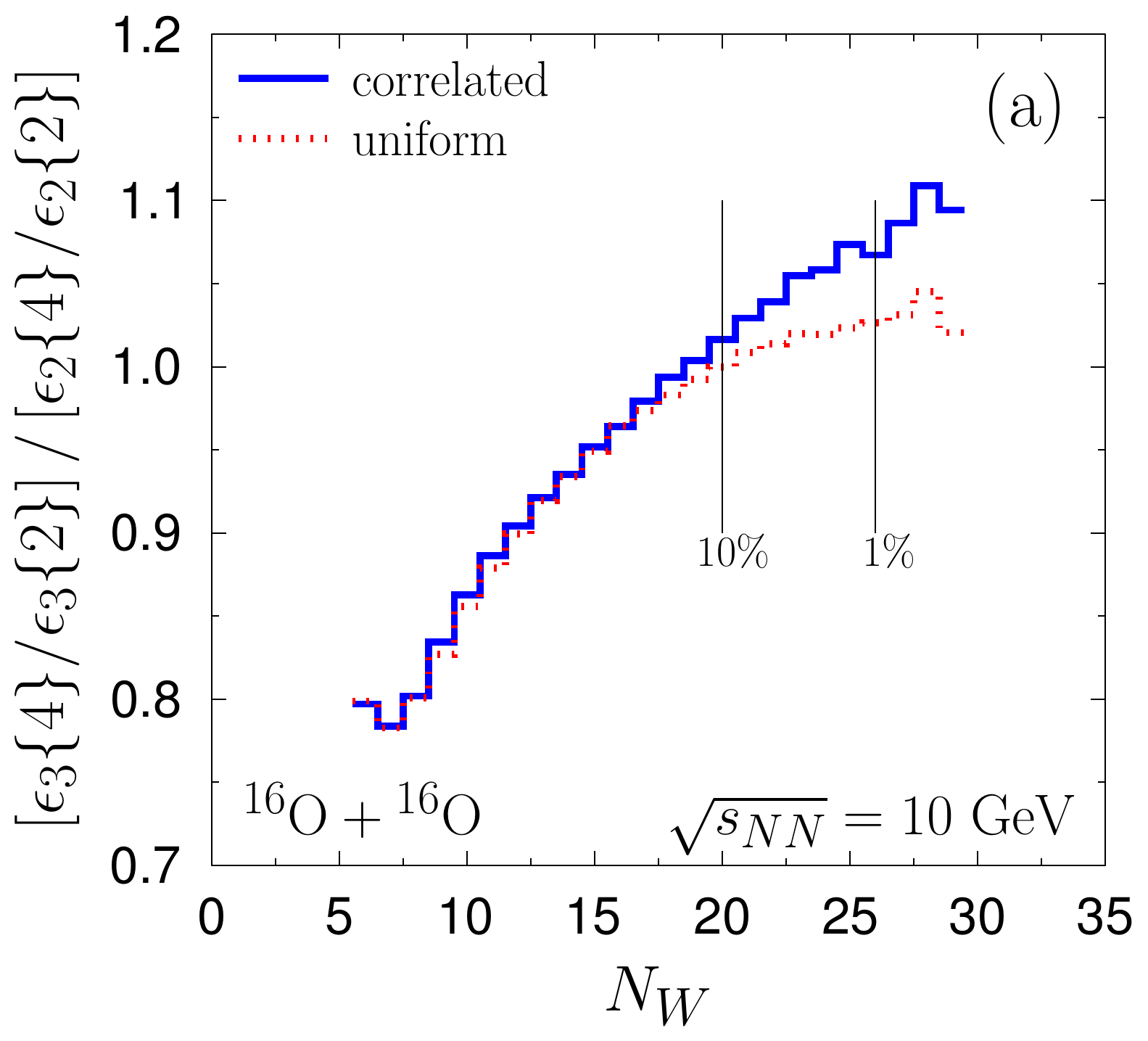} \\
\includegraphics[width=0.4 \textwidth]{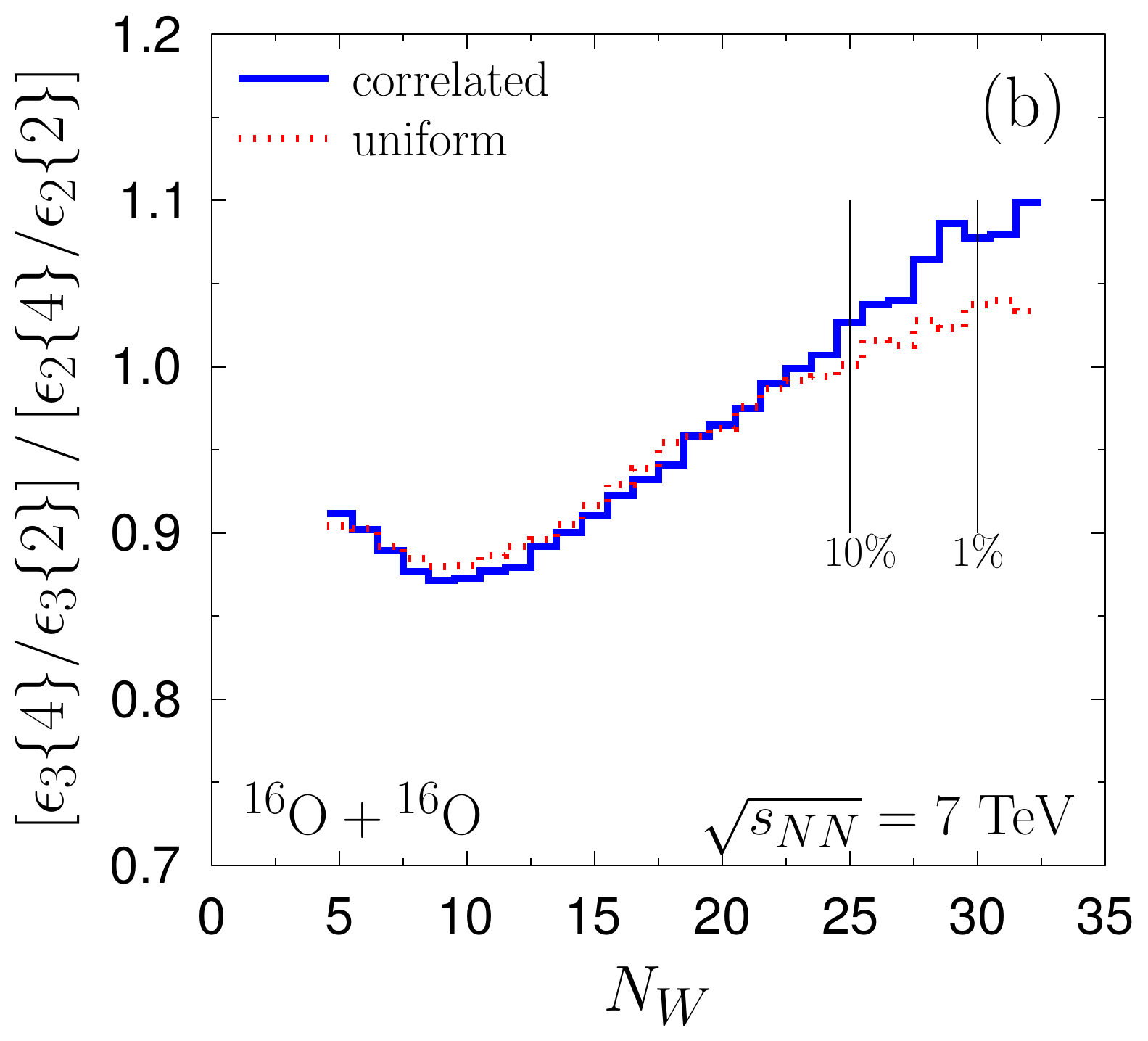} 
\end{center}
\vspace{-5mm}
\caption{Double eccentricity ratio for ${}^{16}{\rm O}+{}^{16}{\rm O}$ collisions at $\sqrt{s_{NN}}=10$~GeV (a) and $\sqrt{s_{NN}}=7$~TeV (b)
plotted as functions of the number of wounded nucleons, $N_w$. \label{fig:double_ratio_OO}}
\end{figure}

\begin{figure}
\begin{center}
\includegraphics[width=0.4 \textwidth]{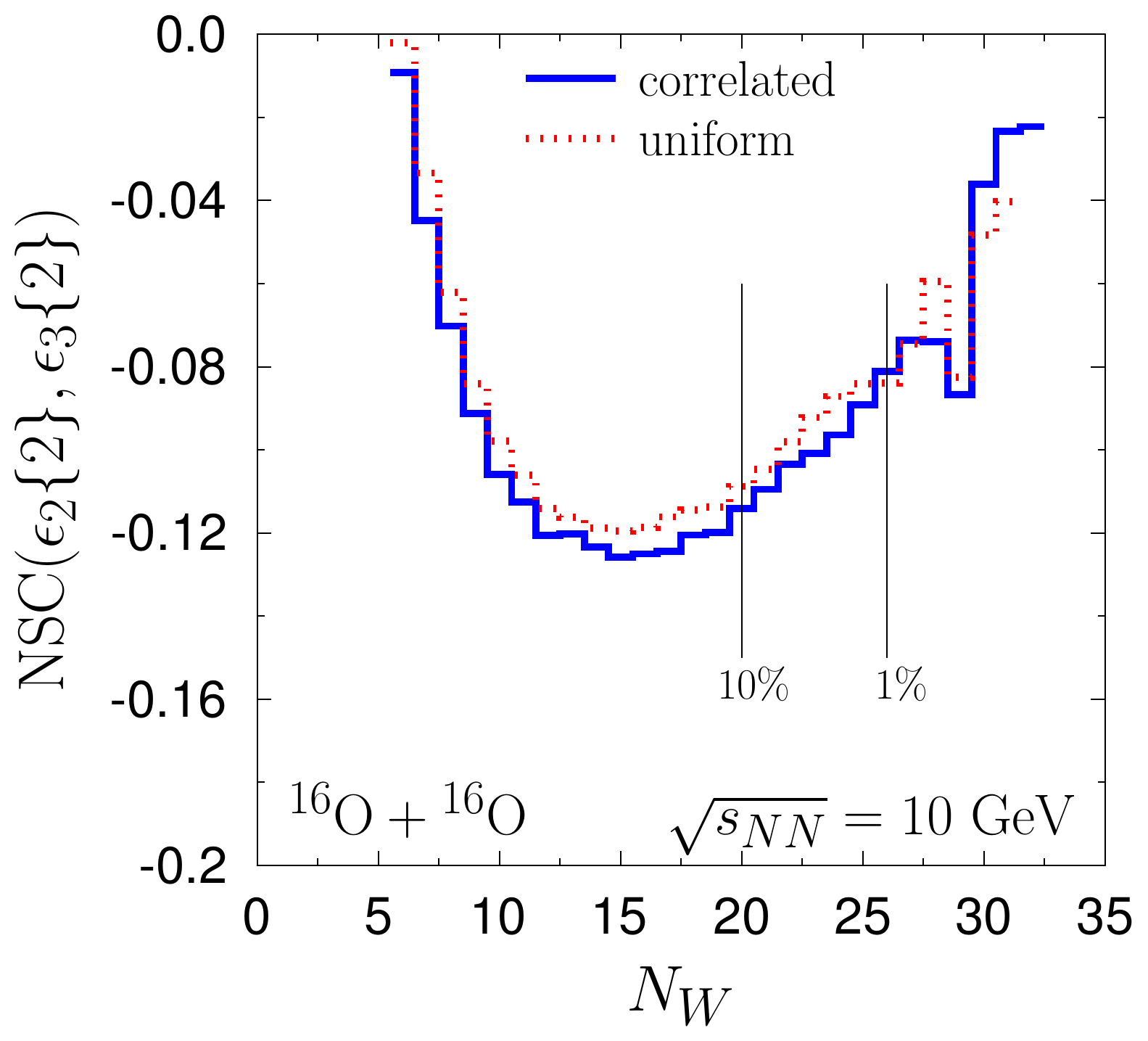} 
\end{center}
\vspace{-5mm}
\caption{Normalized symmetric cumulant for ${}^{16}{\rm O}+{}^{16}{\rm O}$ collisions  at  $\sqrt{s_{NN}}=10$~GeV for the correlated and uniform 
nuclear distributions, plotted as functions of the number of wounded nucleons, $N_w$. \label{fig:sc_OO_comp}}
\end{figure}

\begin{figure}
\begin{center}
\includegraphics[width=0.4 \textwidth]{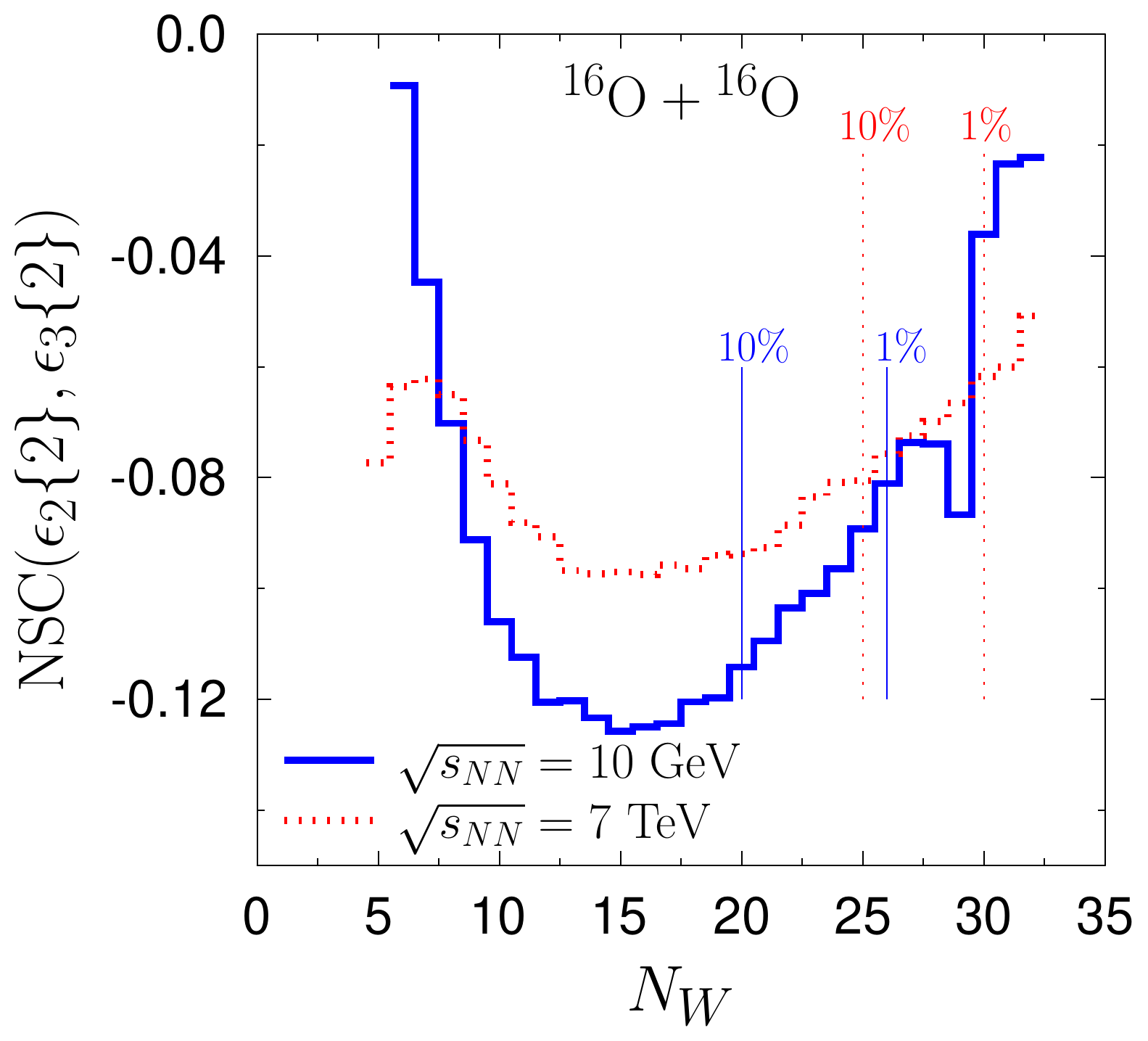} 
\end{center}
\vspace{-5mm}
\caption{Normalized symmetric cumulant for ${}^{16}{\rm O}+{}^{16}{\rm O}$ collisions at $\sqrt{s_{NN}}=10$~GeV and $\sqrt{s_{NN}}=7$~TeV
plotted as functions of the number of wounded nucleons, $N_w$.  Correlated nuclear distributions. \label{fig:sc_OO}}
\end{figure}

We note that the statistics of 6000 ${}^{16}{\rm O}$ configurations allows us to construct $\sim$18~M different collision events
(not counting the random rotation of the nuclei and the change of the impact parameter), which is 
statistically more than sufficient for our studies.
In the following we consider two collision energies for  ${}^{16}{\rm O}$+${}^{16}{\rm O}$:  
$\sqrt{s_{NN}}=10$~GeV, which is accessible in the beam energy scan at RHIC or at SPS, and 
$\sqrt{s_{NN}}=7$~TeV, which has been studied at the LHC. The difference between these energies is in the value of the NN inelastic cross 
section (which grows from 31~mb up to 71~mb between these energies) and in the NN inelasticity profile~\cite{Bozek:2019wyr}. 

First, we discuss the distribution of the number of wounded nucleons, $N_{\rm w}$, in ``minimum bias'' events, that is, for random unconstrained 
values of the ${}^{16}{\rm O}$+${}^{16}{\rm O}$ impact parameter. The results for the two collision energies are compared in Fig.~\ref{fig:nw}. 
We use the logarithmic scale, as typically done in experimental analyses. We note that at the higher energy the distribution is more flat at higher $N_{\rm w}$
(with the obvious limit at $N_{\rm w}=32$), since the higher value of the inelastic cross section makes it easier to wound more nucleons. The vertical 
lines indicate the corresponding centralities, obtained as quantiles of the distribution of $N_{\rm w}$.

\begin{figure}
\begin{center}
\includegraphics[width=0.4 \textwidth]{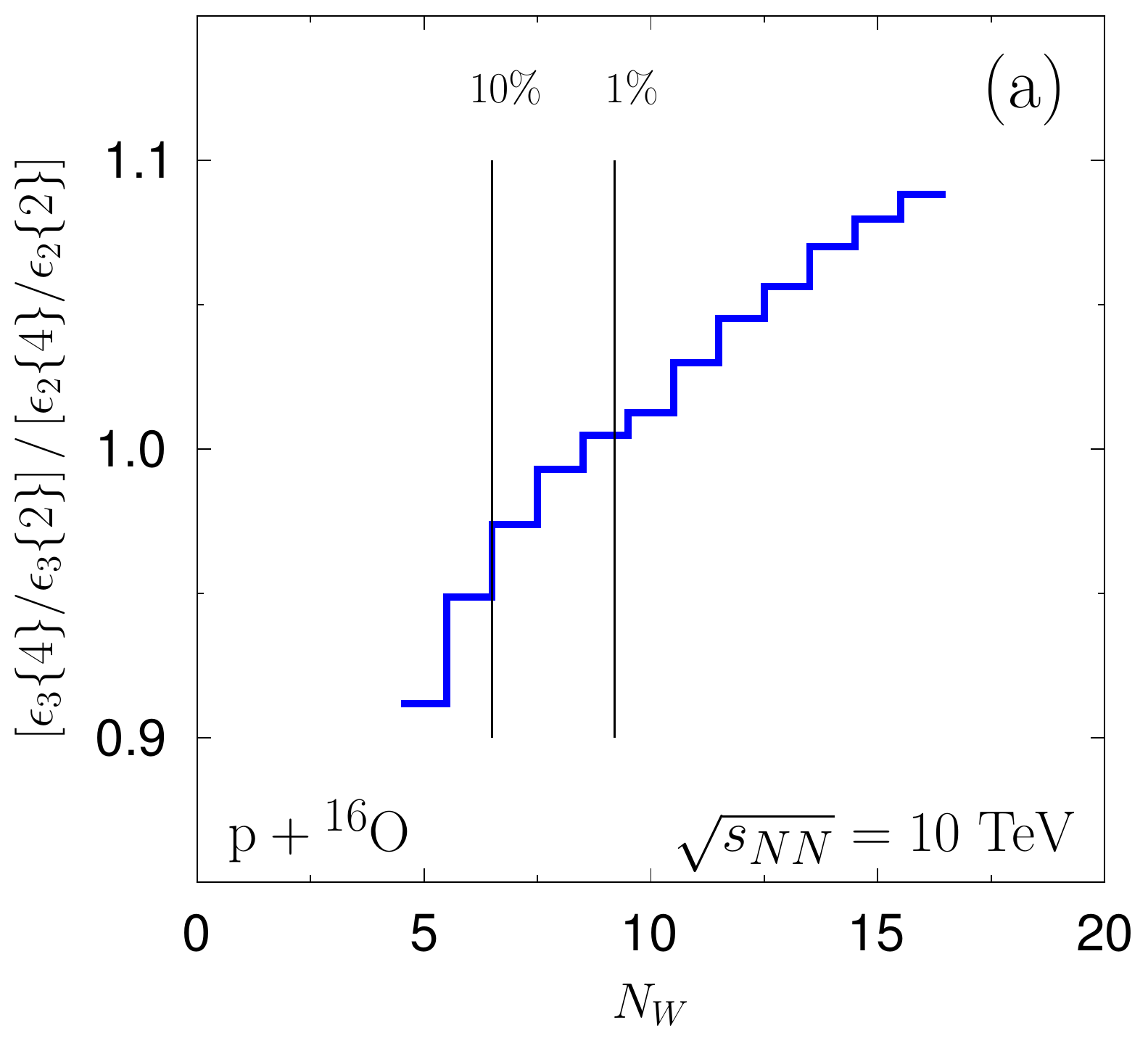} \\
\includegraphics[width=0.4 \textwidth]{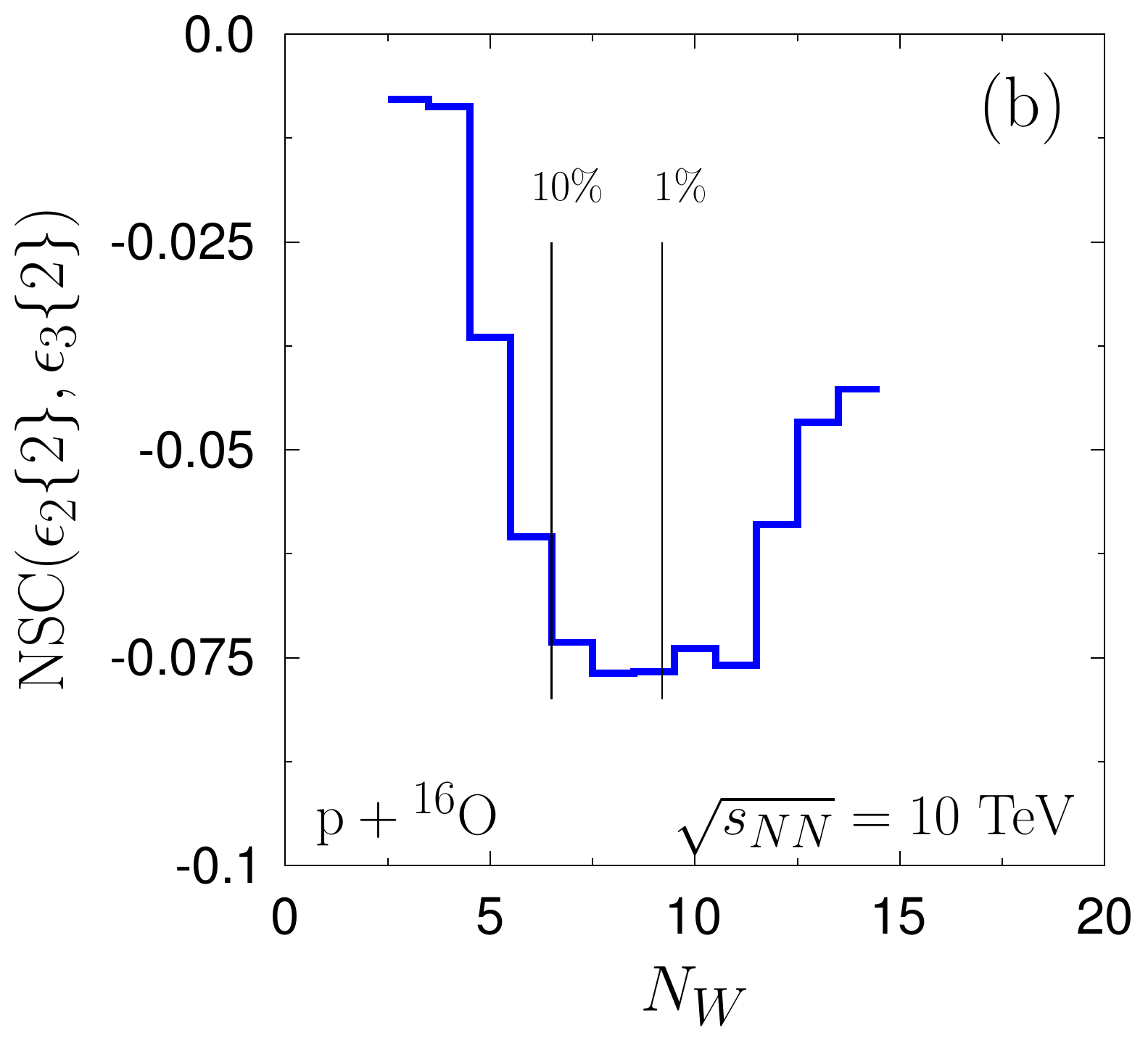} 
\end{center}
\vspace{-5mm}
\caption{Double eccentricity ratio (a) and the normalized symmetric cumulant (b) for p+${}^{16}{\rm O}$ collisions 
at $\sqrt{s_{NN}}=10$~TeV. \label{fig:pO}}
\end{figure}

\begin{figure}
\begin{center}
\includegraphics[width=0.4 \textwidth]{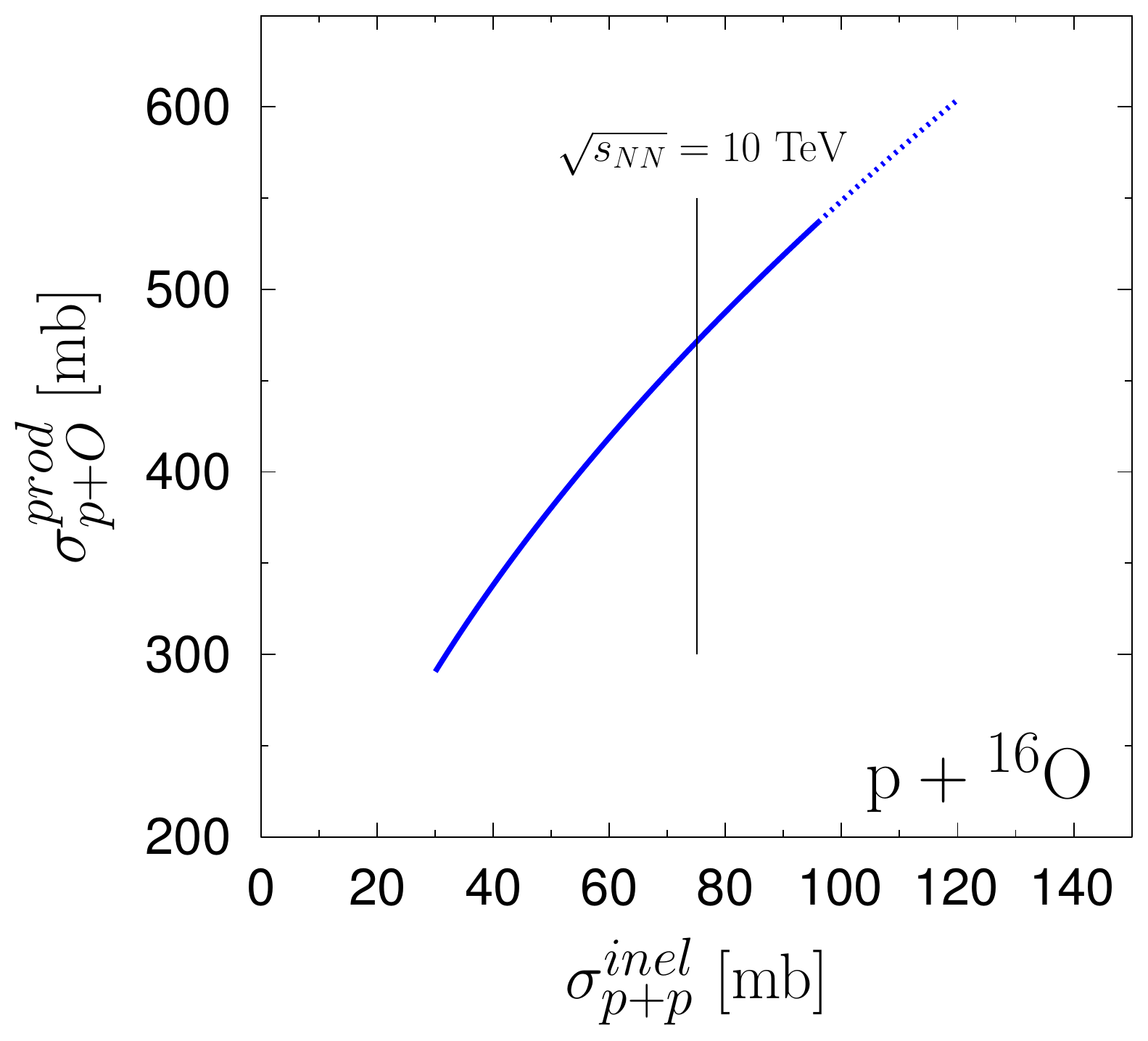} 
\end{center}
\vspace{-5mm}
\caption{Glauber Monte Carlo predictions for the total inelastic
p+${}^{16}{\rm O}$ cross section, plotted as a function of the NN inelastic cross section. 
The planned collision energy of $\sqrt{s_{NN}}=10$~TeV is indicated with a vertical line. The solid line presents results within the 
collision energy range implemented in {\tt GLISSANDO 3}, whereas the dashed line is an extrapolation. \label{fig:xs}}
\end{figure}

\section{Flow signatures in  ${}^{16}{\rm O}+{}^{16}{\rm O}$ collisions \label{sec:flow}}

This section contains the key results in view of the considered  future ${}^{16}{\rm O}+{}^{16}{\rm O}$ experiments.
As this work is based of investigation of the initial condition obtained from the Glauber approach, we are 
going to focus on observables which are to a large degree independent of the hydrodynamic or transport expansion. 
This methodology is based on the shape-flow transmutation feature, appearing in hydrodynamic~\cite{Heinz:2013th,Gale:2013da,Jeon:2016uym} 
or transport simulations~\cite{Lin:2004en}, whereby the deformation of the initial transverse shape of the fireball leads to harmonic 
flow of the hadrons emitted at the end of the evolution. Moreover, the effect is manifest in an approximate proportionality of the flow 
coefficients $v_n$ to the initial eccentricities  $\epsilon_n$, holding for $n=2$ and $3$ and for sufficiently central collisions:
\begin{equation}
v_n \simeq \kappa_n \epsilon_n,  \;\;\; (n=2, 3).
\label{eq:linear}
\end{equation} 
The response coefficients, $\kappa_n$, depend on such features of the colliding system as masses of the projectiles, centrality class, or the collision energy, 
but are to a good approximation~\cite{Gardim:2011xv,Niemi:2012aj} independent of the eccentricities, hence  linearity follows.
For higher rank $n$, as well as for collisions with few participants, 
non-linear effects~\cite{Gardim:2011xv,Giacalone:2017uqx} spoil proportionality (\ref{eq:linear}), hence care is needed in its application.

In practical terms, Eq.~(\ref{eq:linear}) means that one can form ratios of flow observables where the response coefficient $\kappa_n$ cancels out, for 
instance for the cumulant coefficients~\cite{Borghini:2001vi,Bozek:2014cva,Ma:2016hkg,Giacalone:2017uqx} obtained with $k_1$ and $k_2$ particles, 
\begin{eqnarray}
\frac{v_n\{k_1\}}{v_n\{k_2\}} \simeq \frac{\epsilon_n\{k_1\}}{\epsilon_n\{k_2\}},  \;\;\; (n=2, 3), \label{eq:32}
\end{eqnarray}
or the normalized symmetric cumulants~\cite{Bilandzic:2013kga,Mordasini:2019hut} obtained with $k$ particles
\begin{eqnarray}
{\rm NSC}(v_2\{k\},v_3\{k\}) \simeq {\rm NSC}(\epsilon_2\{k\},\epsilon_3\{k\}), \label{eq:sc}
\end{eqnarray}
where
\begin{eqnarray}
{\rm NSC}(a,b)=\frac{\langle a^2 b^2 \rangle }{\langle a^2 \rangle \langle b^2 \rangle }. \label{eq:defnsc}
\end{eqnarray}
In addition to ${\rm NSC}(\epsilon_2\{2\},\epsilon_3\{2\})$, in following we frequently use the ``double eccentricity ratio''
\begin{eqnarray}
\frac{\epsilon_3\{4\}/\epsilon_3\{2\}}{\epsilon_2\{4\}/\epsilon_2\{2\}},
\end{eqnarray}
used in~\cite{Bozek:2014cva}  as a possible probe of the $\alpha$ clusterization in ${}^{12}$C+Au collisions.

We begin the presentation of our Glauber model results for ${}^{16}{\rm O}+{}^{16}{\rm O}$ collisions 
with the ellipticity and triangularity of the fireball obtained with two- and four-particle cumulants, where specifically 
\begin{eqnarray}
&& \epsilon_n\{2\}^2 = \langle \epsilon_n^2 \rangle, \label{eq:ecc} \\ 
&& \epsilon_n\{4\}^4 = 2 \langle \epsilon_n^2 \rangle^2 - \langle \epsilon_n^4 \rangle . \nonumber
\end{eqnarray}
These observables, of course, are not independent of the hydrodynamic response, yet it is worth to 
have a look at them, as they quantify the shape of the fireball and its fluctuations. 

In Figs.~\ref{fig:eps_2} and~\ref{fig:eps_4} we can see the behavior of
the eccentricities. The ellipticity decreases, as expected, with the increasing number of participants, which is the result 
of the geometry (the fireball is less deformed for the central collisions than for the peripheral collisions). Triangularity, due 
entirely to fluctuations, at the lower 
collision energy exhibits a non-monotonic behavior, with maximum around $N_{\rm W}=12$. 

Passing from the two- to four-particle cumulants reduces the eccentricities, as expected from 
the considerations of fluctuations~\cite{Borghini:2001vi}, which increase $\epsilon_n\{2\}$
compared to $\epsilon_n\{k\}$, $k=4,6,\dots$, which are approximately equal~\cite{Bzdak:2013rya}.
This reduction effect can be noted from Fig.~\ref{fig:eps_rat} which displays the ratio of eccentricities 
from four- and two-particle cumulants. 

In Fig.~\ref{fig:eps_22_comp} we investigate the effects of nuclear correlations present in the used distributions 
from~\cite{Lonardoni:2017egu} for $\epsilon_2\{2\}$ (solid line), comparing it to the case where the 
correlations are removed by the mixing technique described in Sec. II (dashed line). We note that the 
difference is small, at the level of a few percent, with the presence of correlations raising the value of  $\epsilon_2\{2\}$ for peripheral collisions, and decreasing it for 
central collisions. Similar size effects appear for other eccentricity coefficients. 
In Fig.~ \ref{fig:double_ratio_OO} we present an analogous study of the double eccentricity ratio. 
We note that the effect of correlations shows up only for the most central events ($c<10\%$) and reaches of a relative size of about 5\% at $c=1\%$.
An analogous study of the normalized symmetric cumulant shown in Fig.~\ref{fig:sc_OO_comp} leads to a similar conclusion. 
We note that ${\rm NSC}(\epsilon_2\{2\},\epsilon_3\{2\})$ remains negative for all centralities (values of $N_{\rm W}$) and exhibits a non-monotonic behavior, both 
at low and high collision energies, as can be inferred from Fig.~\ref{fig:sc_OO}.

\begin{figure}
\begin{center}
\includegraphics[width=0.4 \textwidth]{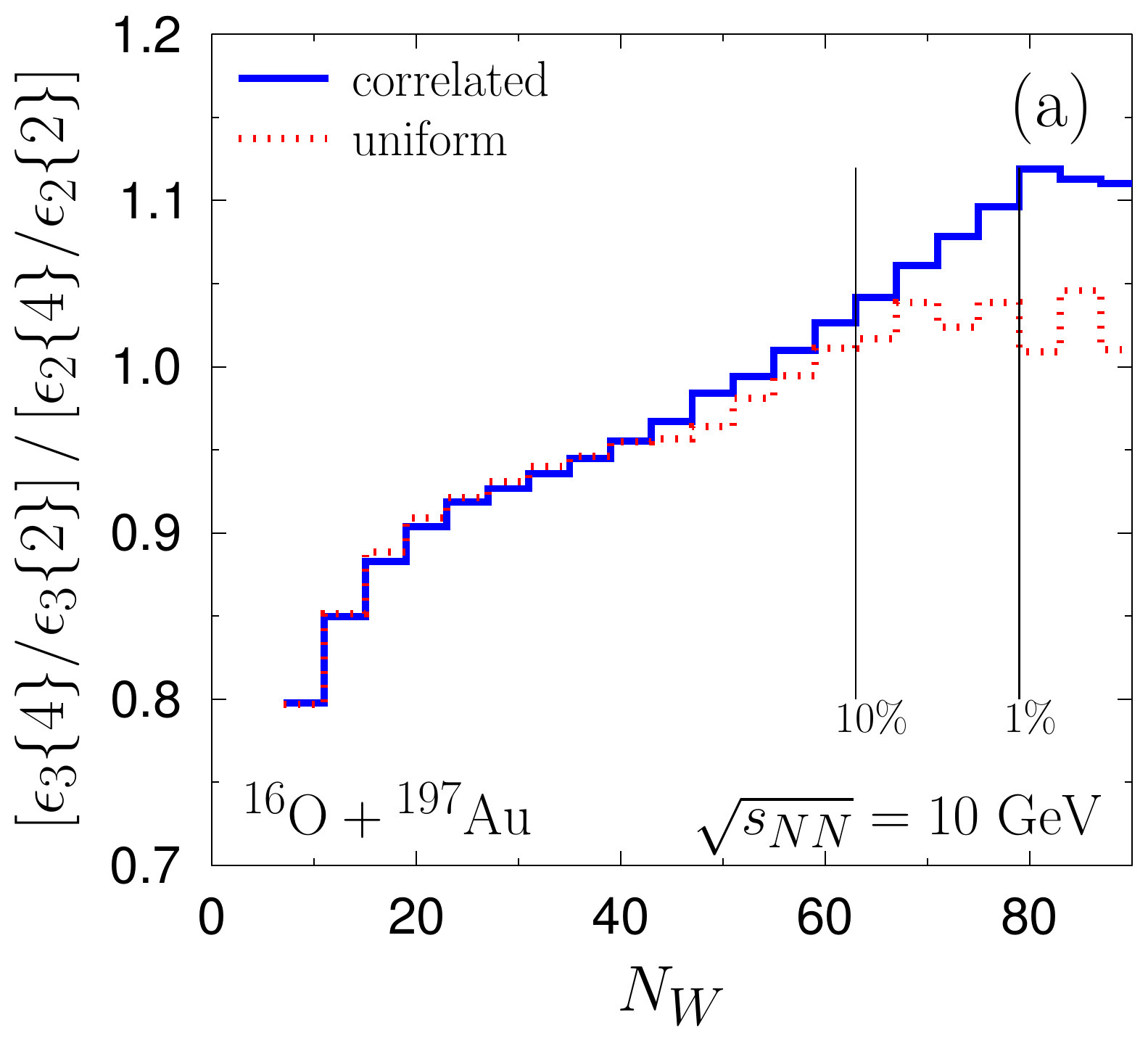} \\
\includegraphics[width=0.4 \textwidth]{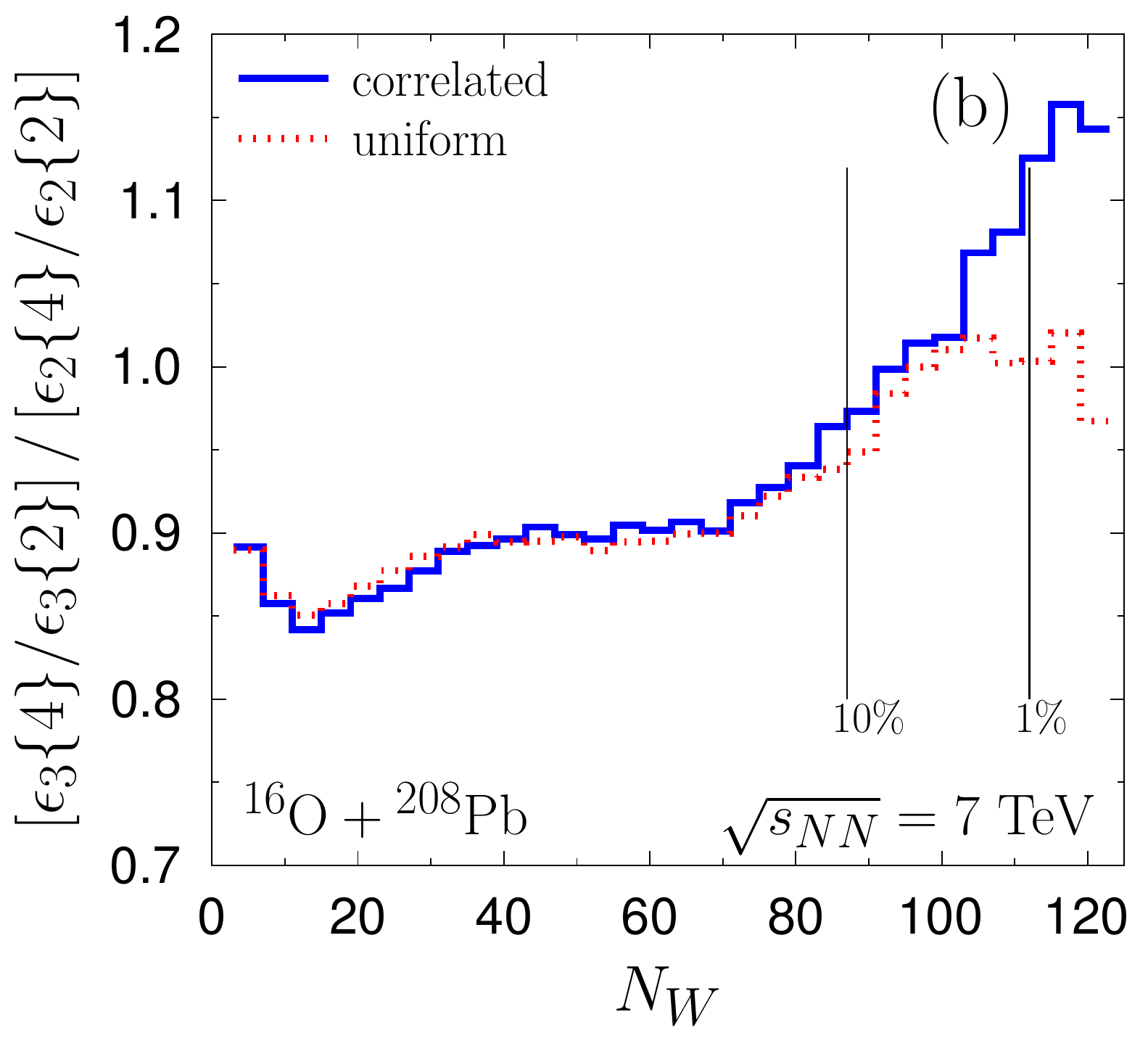} 
\end{center}
\vspace{-5mm}
\caption{Double eccentricity ratio for ${}^{16}{\rm O}+{}^{197}{\rm Au}$ collisions at $\sqrt{s_{NN}}=10$~GeV (a)  and
${}^{16}{\rm O}+{}^{208}{\rm Pb}$ at $\sqrt{s_{NN}}=7$~TeV (b), plotted as a function of $N_{\rm w}$ for correlated and uniform
${}^{16}{\rm O}$ distributions. \label{fig:double_ratio_OAu}}
\end{figure}

\begin{figure}
\begin{center}
\includegraphics[width=0.4 \textwidth]{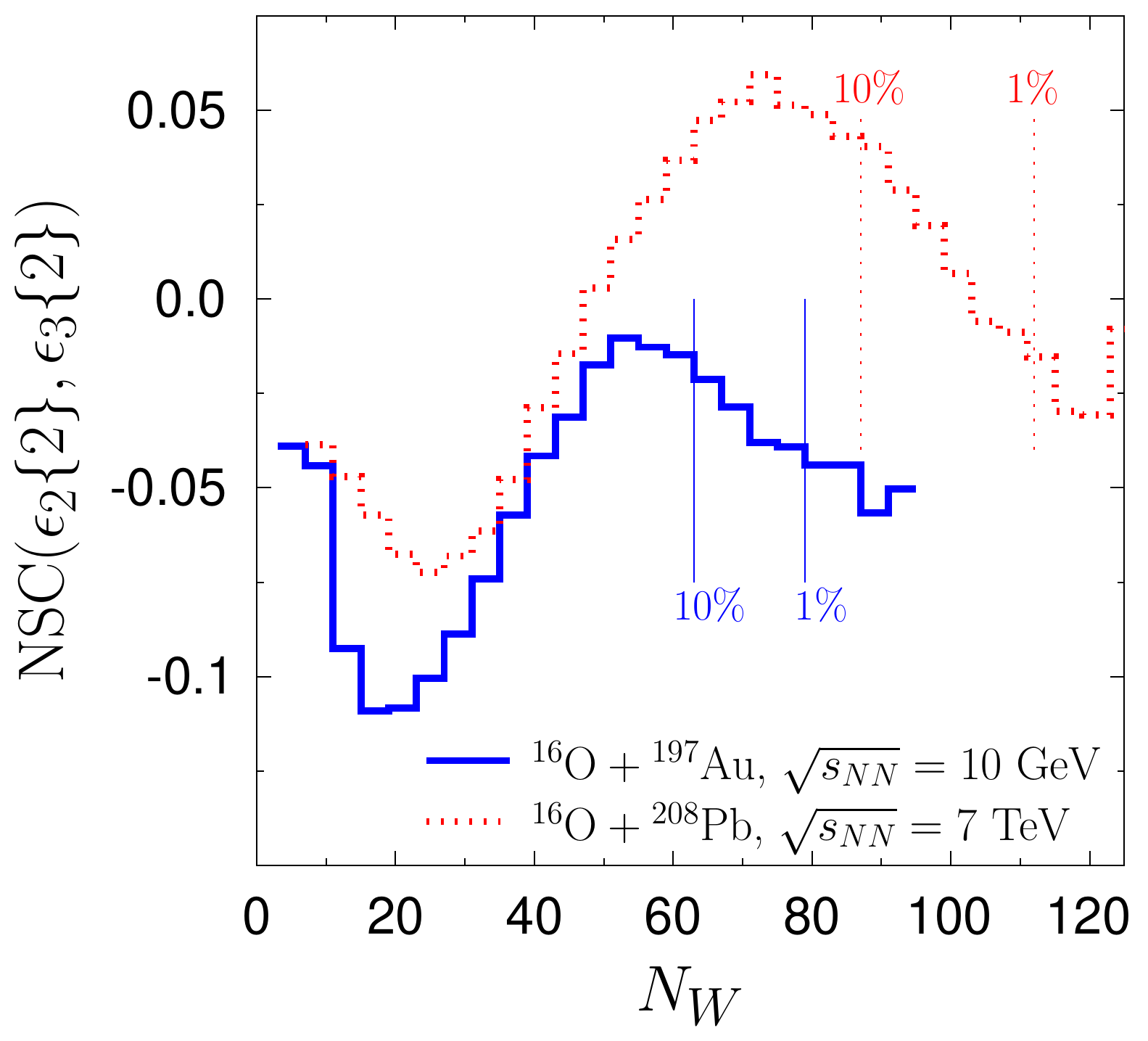} 
\end{center}
\vspace{-5mm}
\caption{Same as in Fig.~\ref{fig:double_ratio_OAu} but for the normalized symmetric cumulants 
(correlated ${}^{16}{\rm O}$ distributions only). \label{fig:sc_OAu}}
\end{figure}

\begin{figure*}
\begin{center}
\includegraphics[width=0.4 \textwidth]{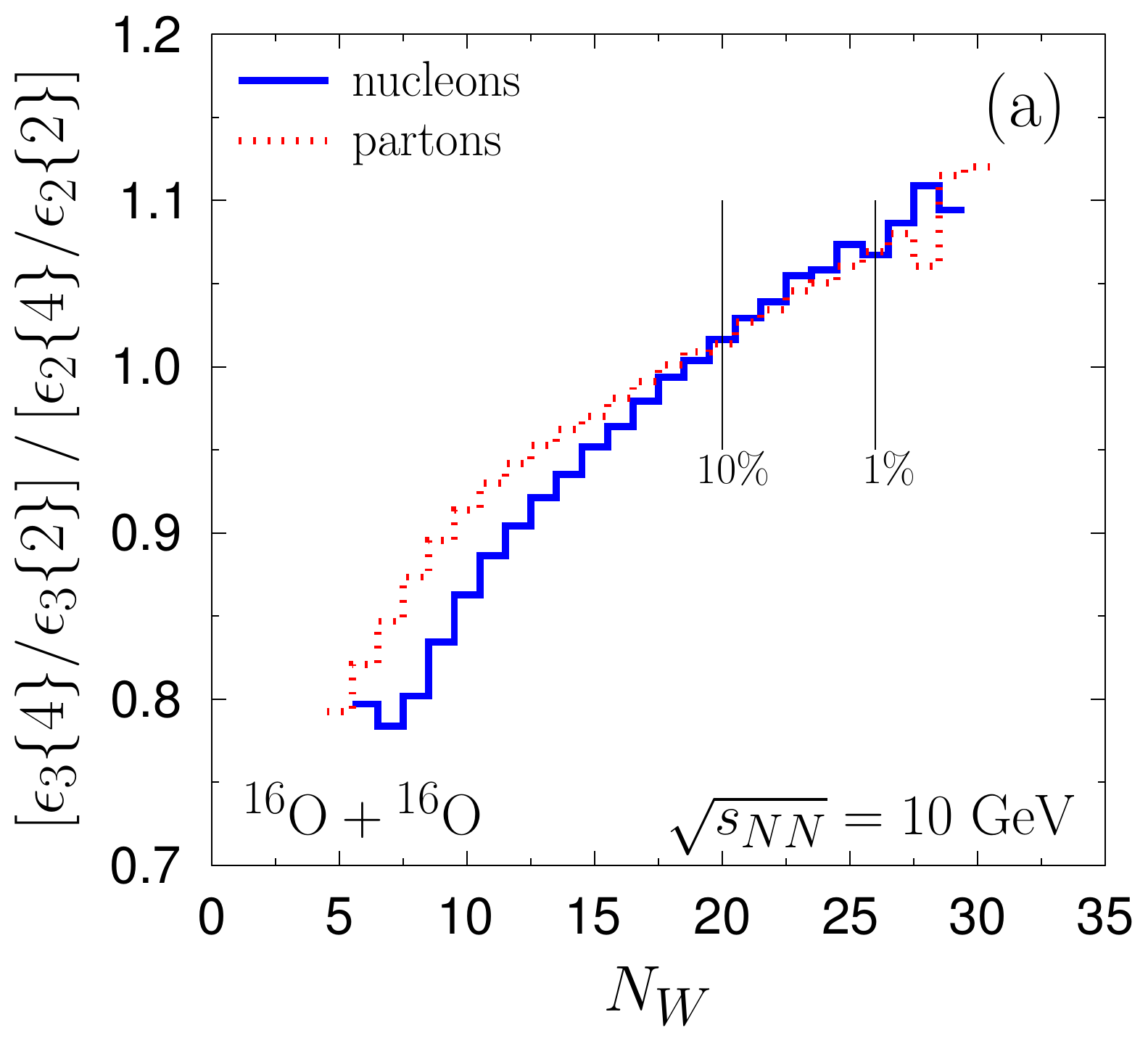} 
\includegraphics[width=0.4 \textwidth]{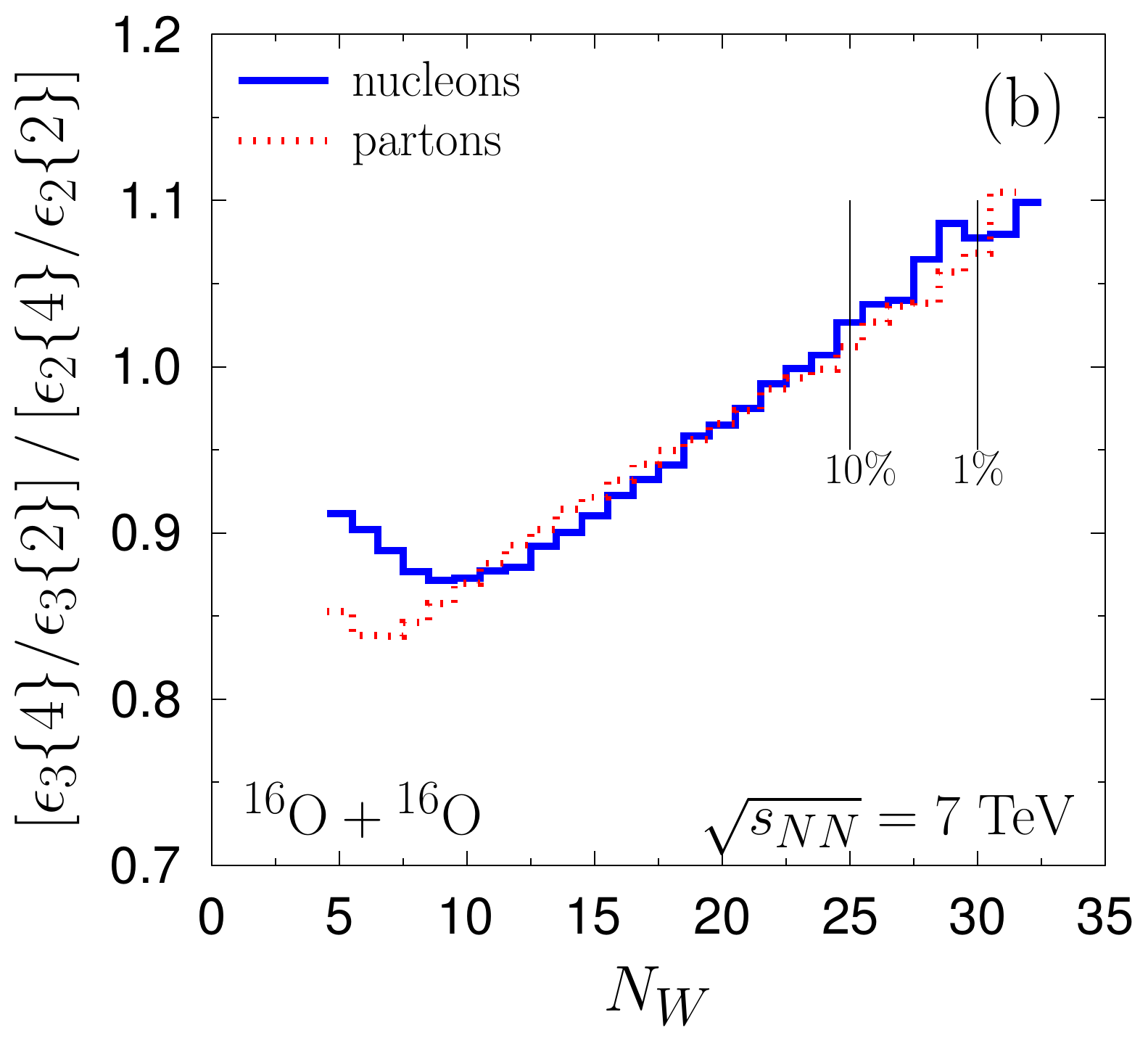} 
\end{center}
\vspace{-5mm}
\caption{Comparison of the wounded nucleon and wounded parton model for the
double eccentricity ratio for $\sqrt{s_{NN}}=10$~GeV (a) and $\sqrt{s_{NN}}=7$~TeV (b) (correlated ${}^{16}{\rm O}$ distributions only).} 
\label{fig:double_ratio_OO_qn}
\end{figure*} 

\begin{figure*}
\begin{center}
\includegraphics[width=0.4 \textwidth]{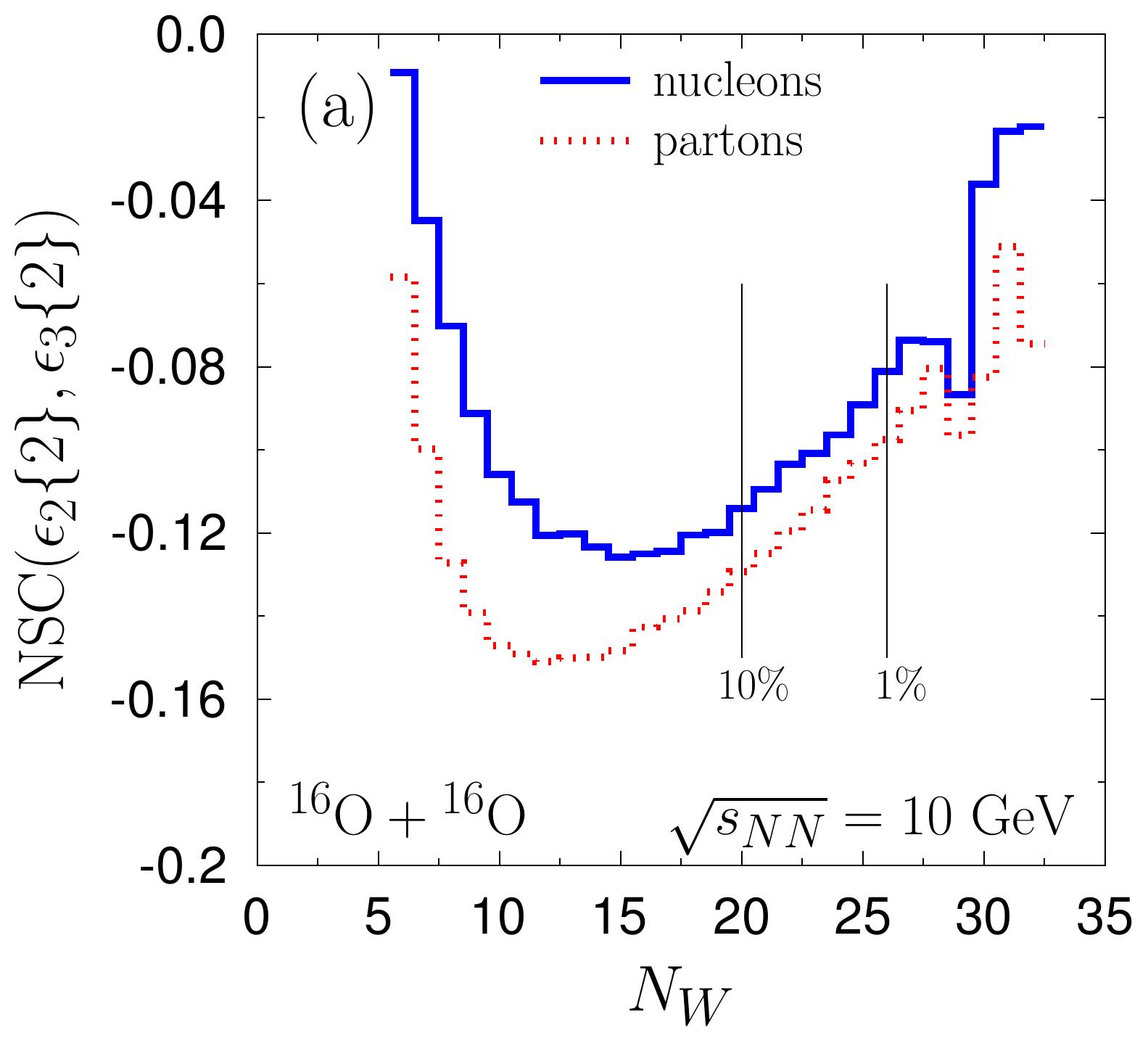} 
\includegraphics[width=0.4 \textwidth]{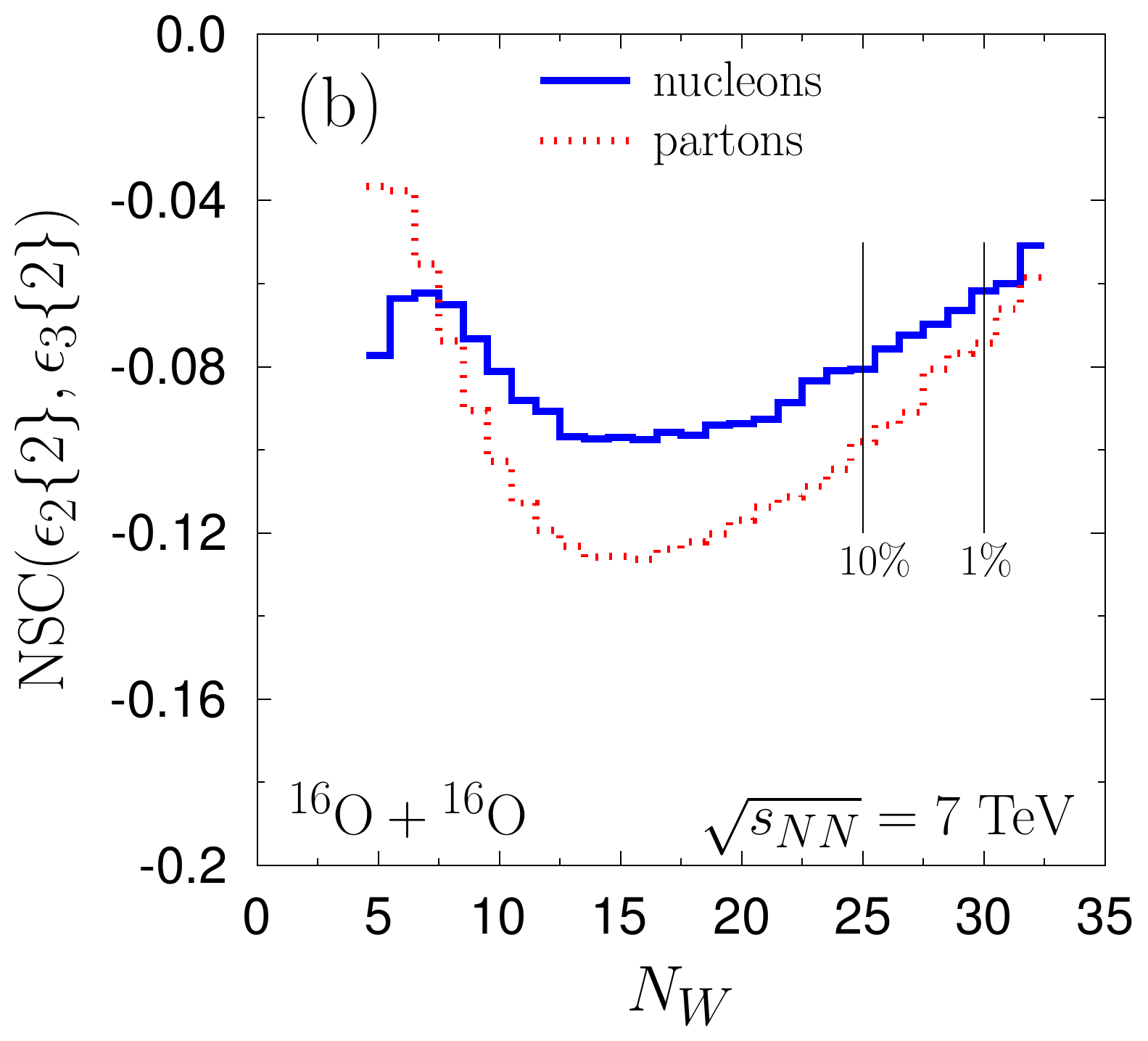} 
\end{center}
\vspace{-5mm}
\caption{Same as in Fig.~\ref{fig:double_ratio_OO_qn} but for the  normalized symmetric cumulant.} 
\label{fig:sc_OO_qn}
\end{figure*}

\begin{figure}
\begin{center}
\includegraphics[width=0.4 \textwidth]{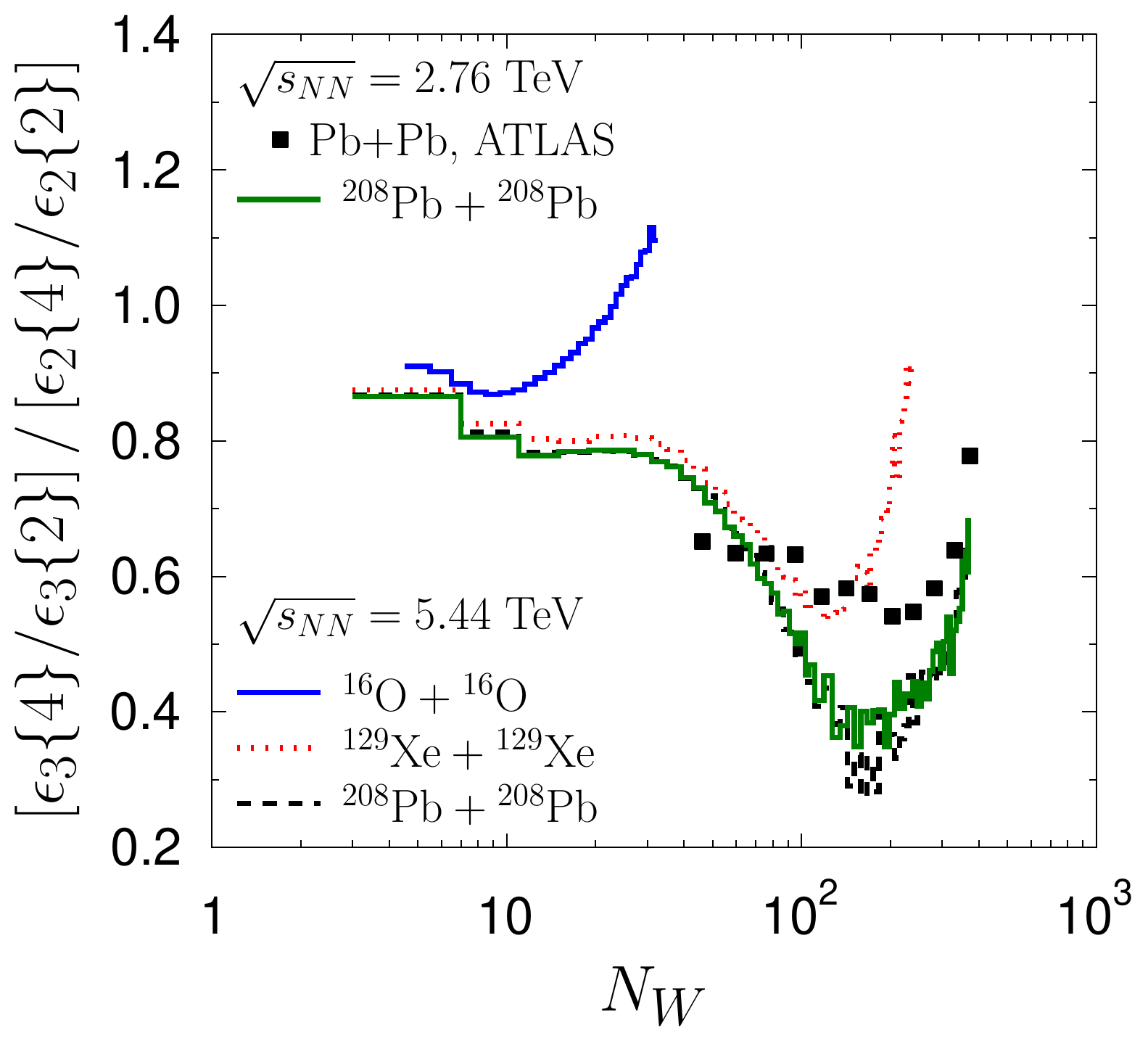} 
\end{center}
\vspace{-5mm}
\caption{Comparison of the Glauber model predictions for the 
double eccentricity ratio in O-O, Xe-Xe, and Pb-Pb collisions at  $\sqrt{s_{NN}}=2.76$~TeV and  $\sqrt{s_{NN}}=5.44$~TeV. 
The experimental data come from \cite{Aad:2014vba}. \label{fig:double_Xe}}
\end{figure}

\begin{figure}
\begin{center}
\includegraphics[width=0.4 \textwidth]{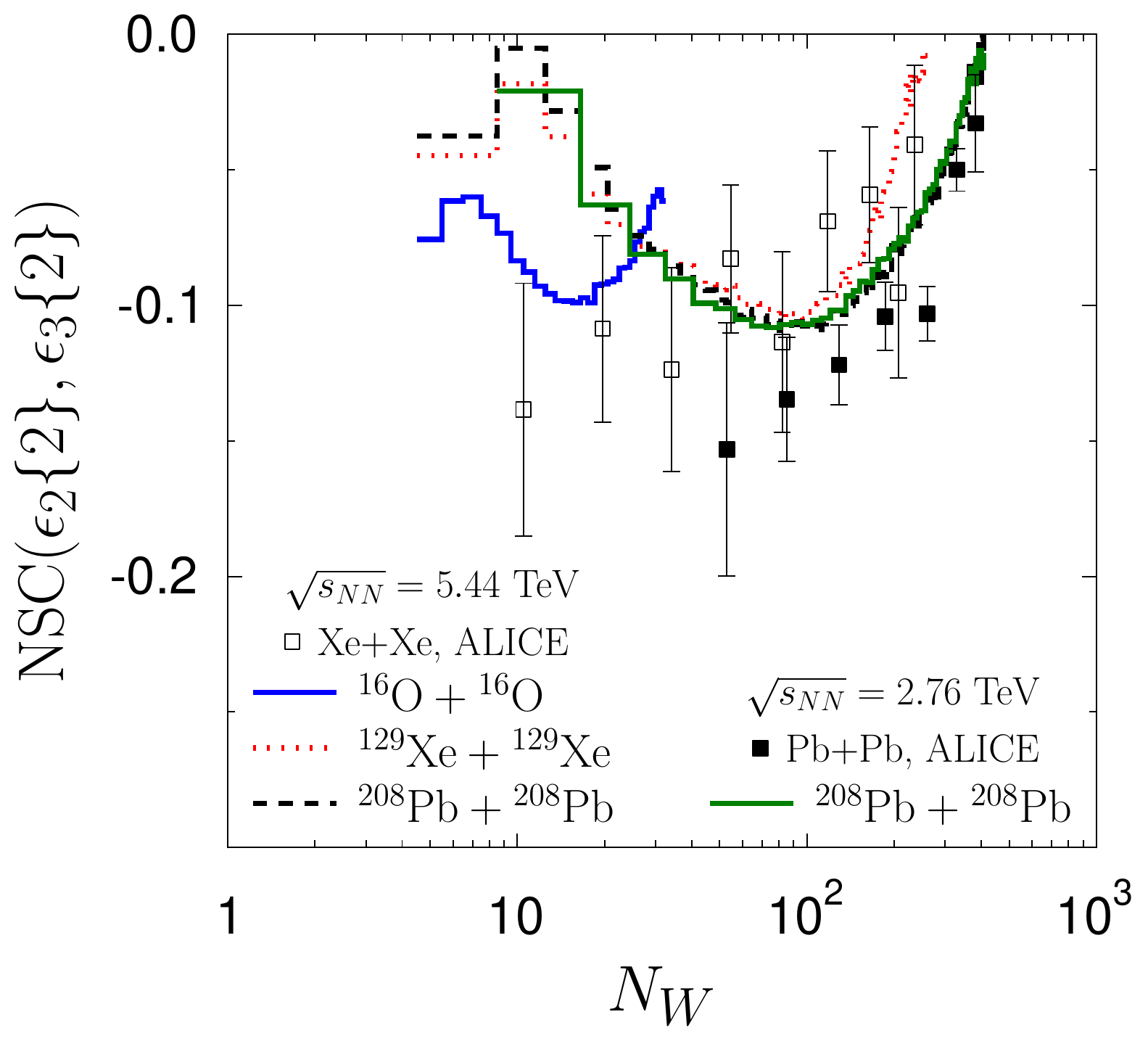} 
\end{center}
\vspace{-5mm}
\caption{Same as in Fig.~\ref{fig:double_Xe} but for the normalized symmetric cumulant. 
The experimental data come from \cite{Acharya:2019vdf, ALICE:2016kpq}. \label{fig:sc_Xe}}
\end{figure}

\section{${\rm p}$+${}^{16}{\rm O}$ collisions \label{second}}

The studies of p+${}^{16}{\rm O}$ reactions at the planned $\sqrt{s_{NN}}=10$~TeV collision energy~\cite{demb}
correspond to interactions in air showers at the proton LAB energy of 50~PeV. In Fig.~\ref{fig:pO} we show our predictions 
for the double eccentricity ratio and the normalized symmetric cumulant. We note that the behavior is qualitatively similar to the 
case of ${}^{16}{\rm O}+{}^{16}{\rm O}$ from Fig.~\ref{fig:sc_OO}.

In Fig.~\ref{fig:xs} we show a quantity relevant to cosmic air-shower considerations, namely, the 
p+${}^{16}{\rm O}$ production cross section, plotted against the NN inelastic cross section, which depends on the collision energy. 
The solid line in the figure presents results within the 
collision energy range $\sqrt{s_{NN}}=5$~GeV -- 57~TeV implemented in {\tt GLISSANDO 3}, whereas the dashed line is an extrapolation made according to the 
fit formula 
\begin{eqnarray}
\sigma^{\rm prod}_{p+O}=(48.3~{\rm mb})\cdot \left(\sigma^{\rm inel}_{p+p}/{\rm mb}\right)^{0.52} .
\label{eq:sig_pO}
\end{eqnarray}

In Ref.~\cite{Collaboration:2012wt}, the Pierre Auger Collaboration obtained for $\sqrt{s_{NN}}= 57$~TeV 
the value $\sigma^{\rm prod}_{p+air}=505^{+28}_{-36}$~mb with the corresponding inelastic proton-proton cross section
$\sigma^{\rm inel}_{p+p}=92^{+9}_{-11}$~mb. From our fit with
Eq.~(\ref{eq:sig_pO}) to the \mbox{${\rm p}+{}^{16}{\rm O}$} {\tt GLISSANDO~3} simulations, we get
$\sigma^{\rm prod}_{p+air}\left(92~ {\rm mb}\right)\simeq 507$~mb, in a good agreement with~\cite{Collaboration:2012wt}.

\section{${}^{16}{\rm O}$ reactions with heavy targets \label{sec:lh}}

In our previous papers~\cite{Broniowski:2013dia,Bozek:2014cva,Rybczynski:2017nrx} we have argued that the heavy-light collisions may reveal 
the cluster correlations in the light projectile. The best case here is probably the ${}^{12}$C nucleus, which is believed to have a significant triangular
$\alpha$-cluster component in the ground-state wave function. 

We note from Fig.~\ref{fig:double_ratio_OAu} that the double eccentricity ratio is sensitive to the nuclear correlations for the most central collisions, with effects of 
a relative size of about 10\% at $c=1\%$. For the uniform case, the curves visibly flatten for the most central collisions, whereas 
with correlations present they continue growing. The behavior is similar at both collision 
energies. 

For the symmetric cumulants shown in Fig.~\ref{fig:sc_OAu} there is only 
some moderate difference between the correlated and uniform  ${}^{16}{\rm O}$ distributions, hence we present only the correlated case.  
We note a characteristic 
non-monotonic behavior with a minimum at low $N_{\rm w}$ and a maximum at  intermediate $N_{\rm w}$.

\section{${}^{16}{\rm O}+{}^{16}{\rm O}$ collisions with wounded quarks \label{sec:quarks}}

In this section we investigate the possible role of nucleon substructure in flow signature of ${}^{16}{\rm O}+{}^{16}{\rm O}$ 
collisions. We compare the predictions of the wounded nucleon model and the wounded parton (wounded quark) 
model~\cite{Bialas:1977en,Bialas:1977xp,Anisovich:1977av,Bialas:1978ze} with 
three constituents, which has turned out successful phenomenologically in explaining the RHIC and LHC 
data~\cite{Eremin:2003qn,KumarNetrakanti:2004ym,Adler:2013aqf,Adare:2015bua,Lacey:2016hqy,Bozek:2016kpf,Zheng:2016nxx,Mitchell:2016jio,Loizides:2016djv}.
The wounded parton picture is implemented in {\tt GLISSANDO~3}~\cite{Bozek:2019wyr} by placing three partons around the center of each nucleon 
with an appropriate exponential distribution. The parton-parton inelasticity profile is adjusted in such a way that the resulting NN inelasticity profile generated in p-p 
collisions matches the phenomenological form discussed in Sec.~\ref{sec:glauber}. 

The comparison of the two models for ${}^{16}{\rm O}+{}^{16}{\rm O}$ is shown in Figs.~\ref{fig:double_ratio_OO_qn} and \ref{fig:sc_OO_qn}
for the double flow ratio and the normalized symmetric cumulant, respectively. We note
that the differences for the double ratio are small, at the level of a few percent, hence the model predictions for these observables 
are robust with respect to the inclusion of the partonic substructure. For the case of the symmetric cumulant the differences are more visible, with a more
prominent minimum occurring for the partonic case.

\section{Comparison of ${}^{16}{\rm O}+{}^{16}{\rm O}$ collisions to heavy-heavy collisions \label{sec:xepb}}

Finally, we compare the predictions for ${}^{16}{\rm O}+{}^{16}{\rm O}$ to the results of the heavy-heavy collisions. We take 
here the collision energy of $\sqrt{s_{NN}}=5.44$~TeV, where the data on Xe+Xe are available \cite{Acharya:2019vdf}, together with  
Pb+Pb at a close collision energy of  $\sqrt{s_{NN}}=2.76$~TeV, where also the data have been collected \cite{ALICE:2016kpq, Aad:2014vba}.

Our results for the double eccentricity ratio are shown in Fig.~\ref{fig:double_Xe}. We note the same pattern in the dependence on $N_w$ 
in all the three reactions, with the minima shifted to the left and upwards with the decreasing projectiles' mass. This feature 
simply reflects the increasing value of $N_{\rm w}$ with the mass.

For the case of the normalized symmetric cumulant presented in Fig.~\ref{fig:sc_Xe}, a similar pattern is observed 
for all considered collision systems. We note a hallmark non-monotonicity, and negative values of the function at all values of $N_{\rm w}$.
The cases for Xe-Xe and Pb-Pb collisions agree reasonably well with the data for the symmetric cumulants corresponding to the harmonic flow.

\section{Conclusions \label{sec:cr}}

We have provided a comprehensive  Glauber Monte Carlo analysis of ultra-relativistic reactions with ${}^{16}{\rm O}$ nuclei, including 
${}^{16}{\rm O}$+${}^{16}{\rm O}$, p+${}^{16}{\rm O}$, and ${}^{16}{\rm O}$ collisions on heavy targets. Although our study is 
limited to the properties of the initial condition, relying on eccentricities evaluated in the model, it bares significance for experimental studies using harmonic flow, 
since we apply specific measures  approximately independent of the hydrodynamic or transport response of the system. We have also studied the case of the 
wounded quark model, which leads to similar predictions as the wounded nucleon model. 

In our analysis we have used correlated nuclear distributions for ${}^{16}{\rm O}$, 
with the conclusion that some characteristic features may be searched for in the most central collisions, for instance for the double eccentricity ratio, where 
the nuclear correlations lead to effects for  most central collisions at the level of 10\%.

Our basic conclusions are that within the applied collective framework no major qualitative differences should be expected from comparisons of 
flow characteristics in ${}^{16}{\rm O}$+${}^{16}{\rm O}$ 
collisions to the case of the earlier-studied heavy-ion collisions, such as Xe+Xe or Pb+Pb. If confirmed experimentally, it would hint to a similar 
collectivity-based mechanism of the fireball evolution across these systems, from small to large. 
Some opportunities would also be offered by ${}^{16}{\rm O}$ collisions on a heavy target, where the internal correlation structure in the light nucleus is expected to be 
of relevance to the harmonic flow characteristics.

\acknowledgments
The numerical simulations were carried out in laboratories created under the project
``Development of research base of specialized laboratories of public universities in Swietokrzyskie region'', POIG 02.2.00-26-023/08, 19 May 2009.
This research  was supported by Polish National Science Centre (NCN) grants 2016/23/B/ST2/00692 (MR) and 2015/19/B/ST2/00937 (WB).

\bibliography{mr_lit,hydr,liter,hydr_0,hydr_2}

\end{document}